\documentclass[preprint2,times,tighten, twocolumn]{aastex62}
\usepackage{amsmath,amssymb,amsthm,textcomp} 
\usepackage{color,soul}
\usepackage{lineno}
\usepackage{subfigure}
\usepackage{graphicx}
\usepackage{rotating}
\usepackage[bookmarks=false]{hyperref}

\usepackage[figure,figure*]{hypcap}
\usepackage{newtxmath} 
\shorttitle{Exo-Io Sodium Clouds }
\shortauthors{Oza et al.}
\begin{document}

\title{Sodium and Potassium Signatures of Volcanic Satellites Orbiting Close-in Gas Giant Exoplanets \\}

\author[0000-0002-1655-0715]{Apurva V. Oza}
\affil{Physikalisches Institut, Universit{\"a}t Bern,  Bern, Switzerland}
\author{Robert E. Johnson}
\affiliation{Engineering Physics, University of Virginia, Charlottesville, VA 22903, USA}
\affiliation{Physics, New York University, 4 Washington Place, NY 10003, USA}
\author{Emmanuel Lellouch}
\affil{LESIA-Observatoire de Paris, CNRS, UPMC Univ. Paris 06, Univ. Denis Diderot, Sorbonne Paris Cite, Meudon, France}
\author[0000-0002-6917-3458]{Carl Schmidt}
\affil{Center for Space Physics, Boston University, Boston, USA}
\author[0000-0001-6720-5519]{Nick Schneider}
\affil{Laboratory for Atmospheric and Space Physics,University of Colorado, Boulder, CO, USA}
\author[0000-0001-9446-6853]{Chenliang Huang}
\affil{Department of Physics and Astronomy, University of Nevada, Las Vegas, NV, USA}
\author{Diana Gamborino}
\affil{Physikalisches Institut, Universit{\"a}t Bern, Bern, Switzerland}
\author{Andrea Gebek}
\affil{Physikalisches Institut, Universit{\"a}t Bern, Bern, Switzerland}
\affil{Departement Physik, Eidgen{\"o}ssische Technische Hochschule Z{\"u}rich, Switzerland}
\author[0000-0001-9003-7699]{Aurelien Wyttenbach}
\affil{Leiden Observatory, Leiden University, Netherlands}
\author{Brice-Olivier Demory}
\affil{Center for Space and Habitability, Universit{\"a}t Bern, Bern, Switzerland}
\author[0000-0002-1013-2811]{Christoph Mordasini}
\affil{Physikalisches Institut, Universit{\"a}t Bern, Bern, Switzerland}
\author{Prabal Saxena}
\affil{NASA, Goddard Space Flight Center, Greenbelt, MD, USA}
\author[0000-0003-2769-2089]{David Dubois}
\affil{NASA, Ames Research Center, Space Science Division, Moffett Field, Ca, USA}
\author{Arielle Moullet}
\affil{NASA, Ames Research Center, Space Science Division, Moffett Field, Ca, USA}
\author{Nicolas Thomas }
\affil{Physikalisches Institut, Universit{\"a}t Bern, Bern, Switzerland}

\email{apurva.oza@space.unibe.ch}

\begin{abstract}
Extrasolar satellites are generally too small to be detected by nominal searches. By analogy to the most active body in the Solar System, Io, we describe how sodium (Na I) and potassium (K I) \textit{gas} could be a signature of the geological activity venting from an otherwise hidden exo-Io. Analyzing $\sim$ a dozen close-in gas giants hosting robust alkaline detections, we show that an Io-sized satellite can be stable against orbital decay below a planetary tidal $\mathcal{Q}_p \lesssim 10^{11}$. This tidal energy is focused into the satellite driving a $\sim 10^{5 \pm 2}$ higher mass loss rate than Io's supply to Jupiter's Na exosphere, based on simple atmospheric loss estimates. The remarkable consequence is that several exo-Io column densities are on average \textit{more than sufficient} to provide the $\sim$ 10$^{10 \pm 1}$ Na cm$^{-2}$  required by the equivalent width of exoplanet transmission spectra. Furthermore, the benchmark observations of both Jupiter's extended ($\sim 1000$ R$_J$) Na exosphere and Jupiter's atmosphere in transmission spectroscopy yield similar Na column densities that are purely exogenic in nature. As a proof of concept, we fit the ``high-altitude" Na at WASP 49-b with an ionization-limited cloud similar to the observed Na profile about Io. Moving forward, we strongly encourage time-dependent ingress and egress monitoring along with spectroscopic searches for other volcanic volatiles.
 
\end{abstract}

\keywords{Planets and Satellites: detection; atmospheres;  physical evolution; magnetic fields; composition; dynamical evolution and stability}
\section{Introduction}\label{intro}
The 1970s discoveries of sodium (Na I) and potassium (K I) clouds at Io (\citet{brown74}; \citet{brownchaffee74}; \citet{trafton75} ) turned out to be a revealing observational signature motivating the tidal dissipation theory developed by \citet{peale79} which predicted extreme volcanic activity on Io even before Voyager 1's first images of the system \citep{morabito79}. This activity was confirmed to be globally extensive by remarkable infrared images by subsequent spacecraft missions, decades of direct imaging monitoring (e.g.  \citet{spencer2000}; \citet{marchis05}; \citet{spencer07}; \citet{imke17}; \citet{skrutskie17}; \citet{dekleer16}), and of course the direct detection of volcanic sulfur species (\citet{lellouch90}; \citet{lellouch96SO}). Na \& K observations at Io encouraged the search for the venting parent molecules. The subsequent discovery of ionized chlorine in the plasma torus \citep{kuppers2000}, suggested the presence of the subsequent direct observations of volcanic salts in the mm/sub-mm, NaCl \citep{lellouch03} and KCl \citep{moullet13}. The strong resonance lines in the optical: Na D1 (5895.92 $\AA$), Na D2 (5889.95 $\AA$); K D1 (7699 $\AA$), K D2  (7665 $\AA$) have therefore been pivotal for astronomers characterizing physical processes in atmospheres on Solar System bodies, starting from the early observations of Na I in Earth's upper atmosphere in 1967 (\citet{naearth}; \citet{huntenwallace67}; \citet{hansondonaldson67}). Thanks to advances in remote and in-situ instrumentation, Na I \& K I have been detected in silicate (i.e. Io; Mercury; Moon) and icy (i.e. Europa; comets) bodies, but never in H/He envelopes such as the giant planet atmospheres of our Solar System. In fact, the origin of Jupiter's Na I exosphere extending to $\sim 1000 R_J$ is Io's volcanism interacting with Jupiter's magnetospheric plasma (e.g. \citet{mendillo90}; \citet{wilson02}; \citet{thomas04}). Table \ref{alkalis} summarizes the spectral observations of the Na \& K alkaline metals at several Solar System bodies. \\

Following the first detection of a component of an extrasolar atmosphere, Na I at HD209458b (\citet{charbonneau02}), \citet{johnsonhuggins06} considered the possible effect of material from an orbiting moon, gas torus, or debris ring on the exoplanet transit spectra. At the time it was thought that satellite orbits around close-in gas giant exoplanets (hot Jupiters) might not be stable, therefore \citet{johnsonhuggins06} suggested that large outgassing clouds of neutrals and/or ions might only be observable for giant exoplanets orbiting at $\gtrsim 0.2$ AU. While larger orbital distances are still safer places to search for a satellite, we confirm that close-in gas giant satellites can be more stable than expected. Furthermore, the rapid orbital periods of these close-in satellites enables an efficient transit search. While the above-mentioned Na I exosphere out to $\sim 1000 R_J$ shrinks to $\sim 1 R_J$ due to the far shorter photoionization lifetime of Na I, recent understanding on orbital stability by \citet{cassidy09} results in a stabilizing stellar tide well within $< 0.2$ AU driving up to six orders of magnitude more heat into the satellite sourcing a density enhancement of $\sim$ 3 - 6 orders of magnitude into the planetary system. This in turn results in Na I clouds ranging from $\sim 10^{10} - 10^{13}$ Na cm$^{-2}$ readily discernable by transmission spectroscopy. In fact, a large number of new transmission spectroscopy observations have detected this range of alkaline column densities which we will evaluate individually in this work. This pursuit was also motivated by the recent evaluation of the uncertainties in the interpretation of the alkaline absorption features at hot Jupiters (\citet{heng15}; \citet{heng16}) which has led to the suggestion described here, that an exogenic \footnote{In the following, we refer to processes intrinsic to the gas giant as endogenic and those external, in our case to the satellite or debris ring, as exogenic.} source from an active satellite might not be unreasonable for certain hot Jupiters. Therefore, we first review recent work describing the range of stable orbits for moons at close-in hot Jupiters.
We then analyze a recently published observation of Jupiter showing the spectral signature of Na I has an external origin. We then describe the mass loss history and influence of an irradiated, active exomoon at a hot Jupiter and, finally, translate these mass loss rates to column densities tentatively confirming that such a signature is consistent with observations at a number of hot Jupiters. Lastly, we provide order of magnitude predictions on additional signatures which could more conclusively confirm the first exo-Io.
\begin{table*}[t]
    \centering
    \hspace*{-1.5cm}
\begin{tabular}{lrrrrrrrr}
\hline
\textbf{Body} &   \textbf{N$_{Na}$} [cm $^{-2}$]  &  \textbf{Na/K} & \textbf{Origin}  \\
\hline
\underline{$^{a}$Comets}	& $\sim 8 \times 10^{11}$ & \textbf{Na/K} = 54 $\pm$ 14  & Photo-desorption/Solar wind sputtering \\ 
\underline{$^{b}$Mercury} & $\sim 8 \times 10^{11}$ &\textbf{Na/K} = 80-190 & Photon-stimulated Desorption    \\
\underline{$^{c}$Io} & $\sim 5 \cdot 10^{9} - 10^{12}$   & \textbf{Na/K} = 10 $\pm$ 3 & Atmospheric Sputtering/Pick-up Ions   \\
 & NaCl/KCl = 5.75 $\pm$ 0.5   \\ 
\underline{$^{d}$Europa} & $\sim 2 \times 10^{10}$ & \textbf{Na/K} = 25 $\pm$ 2 & Surface Sputtering; Cryovolcanism? \\
\underline{$^{e}$Moon} & $\sim 5 \times 10^{10}$ & \textbf{Na/K} = 6 &  Photo-desorption \& Meteors   \\
\underline{$^{f}$Enceladus}  & $\lesssim 10^{9}$ & \textbf{Na/K} $\sim 100$ &  Cryovolcanism    \\
\underline{$^{g}$Earth*} & $\sim 8 \times 10^{11}$& \textbf{Na/K} $\sim 100$  & Micrometeorite ablation  \\
\underline{$^{h}$Jupiter**} & $\sim 7 \times 10^{9}$ & Unknown & Iogenic streams; micrometeorites; impacts  \\
\underline{$^{i}$Sun} & Solar Abundance & \textbf{Na/K} $\sim 16 \pm 2$  & Protosolar Nebula  \\
\end{tabular}
    \caption{\textbf{Alkaline Atmospheres} for silicate and icy bodies measured in-situ and remotely in the Solar System. For each body we quote the zenith column density for Na I, if observed. The origin of Na I \& K I for most bodies is due to space weathering or [cryo]volcanism. We note that Europa and Enceladus likely source Na I from a NaCl-rich water ocean, and therefore Na I could also indicate the presence of water. Direct [cryo]volcanism at both Io and Enceladus is also expected to be NaCl-rich, which subsequently dissociates to Na. The K I observed at Io is thought to be of similar origin to Na I, dissociating from KCl as well. The NaCl/KCl ratio in the atmosphere is a factor of two less than the Na/K ratio in the escaping atmosphere, probing the lower and upper atmosphere of Io respectively.  *Earth's Na I observed in the mesosphere originates from ablation of interplanetary dust particles. **Jupiter's upper atmosphere Na I column density is computed in Section \ref{jupitersNa} of this work. Na/K for the solar abundance is also tabulated $\sim 16 \pm 5$. \\
    $^{a}$ \citet{leblanc08}; \citet{schmidt15} \& \citet{schmidt16}; \citet{schmidt16comet} \\
    \citet{fulle2013}\\
    $^{b}$\citet{sprague97}\\
    \citet{pottermorgan85}; \citet{pottermorgan86} \\
    $^{c}$ \citet{burger01}\\
     \citet{trafton75}; \citet{thomas96}\\
     \citet{lellouch03}; \citet{moullet15}\\
    $^{d}$ \citet{leblanc05}; \citet{leblanc02}; \citet{brown96}\\
     \citet{brown2001}\\
    $^{e}$ \citet{szalay16}; \citet{wilson06} \\
    \citet{pottermorgan88} \\
    $^{f}$ \citet{postberg09}; \citet{schneider09} \\
    $^{g}$  Slipher 1929; \citet{naearth}\\
    \citet{sullivanhunten62}, \citet{sullivanhunten64};  \citet{gardner14}\\ 
    $^{h}$ \citet{montanes}; This Work.\\
    $^{i}$  \citet{asplund2009}\\
    }
    \label{alkalis}
\end{table*}
\section{Tidal Stability of an Exo-Io}\label{tidalstability}
The dynamic stability of extrasolar satellites depends strongly on the uncertain tidal factor: $\mathcal{Q}_{p,s} \propto \dot {E}_{p,s}^{-1}$. Here $\dot{E}_p$ is the energy dissipated by tides into the planetary body (orbital decay) and $\dot{E}_s$ into its orbiting satellite (satellite heating) which is also forced by a third body, the host star. \citet{cassidy09} studied both the orbital decay and heating of satellites by considering the circular restricted three-body problem for a satellite, a hot Jupiter, and a host star. It was shown that if the tidal $\mathcal{Q}$ for the planet, $\mathcal{Q}_p$ is of the order of the equilibrium tide limit, $\mathcal{Q}_p \sim 10^{12}$ as first derived by \citet{goldreich77} and improved by \citet{wu05-1}, even Earth-mass exomoons around hot Jupiters could be tidally-stable on $\sim$ Gyr timescales. Kepler data has not yet detected such exomoons, except for the recent tentative identification of a Uranus-sized candidate Kepler 1625-b at $\sim$ 1 AU (\citet{teacheykipping18}; \citet{kreidberg2019}). The observation of a close-in exomoon would in principle be able to constrain the low tidal $\mathcal{Q}$ values used in the literature (\citet{barnesobrien02}; \citet{weidnerhorne10}) previously set to Jupiter's $\sim 10^{5}$ (\citet{laineytobie05}). As shown by \citet{cassidy09}, and later expanded upon in sections \ref{magnetotidal} \& \ref{destruction}, significant mass loss might have substantially eroded large satellites (e.g. \citet{domingos06}), decreasing their ability to be detected by mass-dependent searches such as transit timing variations (e.g., \citet{agol05}; \citet{kippingexomoons09}). An exomoon search \textit{independent} of the satellite size is therefore needed. 

The semimajor axis of a stable exomoon orbiting a close-in gas giant exoplanet of eccentricities e$_s$ and e$_p$ respectively is narrowly confined, $a_s \lesssim 0.49 a_H (1.0 - 1.0 e_p - 0.27 e_s)$ \citep{domingos06}, at roughly half of the Hill radius, $a_H = a_p (\frac{M_p}{3 M_*})^{1/3}$. These orbital and observational parameters, for a sample of 14 gas giants out to $\sim 10 R_*$ are computed in Table \ref{T2stellar}. For nearly circular orbits, we find that for a$_s \lesssim $ 2.3 R$_p$ the satellite remains stable, where R$_p$ is the exoplanet transit radius. Table \ref{T3planet} gives a$_s$ and a$_H$, along with several other length scales such as R$_{i} = R_p + \delta R_i$ the \textit{apparent} transit radius at line center due to the presence of a species $i$, absorbing at resonance wavelength $\lambda_{i}$ adding an apparent change in radius $\delta R_i$. The observed absorption depth, $\delta_i = \frac{\Delta \mathcal{F}_{\lambda,i}}{\mathcal{F}_{\lambda,i}}$ reveals the equivalent width, W$_{\lambda,i} = \delta_i \Delta \lambda_i$, where $\mathcal{F}_{\lambda, i}$ is the fractional change in the stellar flux at line center, and $\Delta \lambda_i$ is the bandpass of the spectrograph. In Section \ref{columnsection} we will describe how this information can yield the approximate column density of the absorbing species. As the geometry of the absorbing gas is unknown, spherical symmetry is often assumed while the exoplanet transits a star of radius R$_*$, so that the absorption depth can also be indicative of the ratio of the effective areas: the absorbing layer (assumed to be an annulus) to the stellar disk, $\delta_{i} = \frac{R_{i}^2 - R_p^2}{R_*^2} \sim \frac{2 \pi R_p \delta R_{i}}{\pi R_*^2}$. The Roche radius further confines the satellite's orbit,  R$_{roche} =  2.456 \left(\frac{\rho_{p}}{\rho_{s}}\right)^{1/3}$ \citep{chandrasekhar69}, inside which a possible debris disk from a disintegrated satellite could be present as discussed later in Section \ref{magnetotidal}.

\indent Our study of a tidally-locked exoplanet system with planetary orbital period $\tau_{orb} = \frac{2\pi}{\omega_{orb}}$ and $\tau_{s} = \frac{2\pi}{\omega_{s}}$ will result in a moon orbiting faster than the planet's rotation $\omega_{s} > \omega_{orb}$, opposite to the Jupiter-satellite system. When solving the dispersion relation $\omega_{s} = k^2 \omega_{orb}$, \citet{wu05-1}; \citet{wu05-2}; \citet{ogilvielin07}; \citet{ogilvie07}; \citet{ogilvie14} found that $\omega_{s} > \omega_{orb}$, specifically deriving $\tau_s < \tau_{orb}/2$ as an orbital stability limit due to the lack of tidal dissipation waves into the gas giant's convective envelope from a satellite.  The low tidal dissipation $\dot{E}_p$ results in a much higher tidal $\mathcal{Q}_p$ again contrasting the $\omega_{s} < \omega_{orb}$ of the Jupiter system, of low tidal $\mathcal{Q}$ \citep{laineytobie05}. 

\citet{cassidy09} showed that when $\mathcal{Q}_p$ is large enough, orbital decay is slow and the satellite will not be destroyed by tidal decay. That is, the satellite is stable after a time t, if $\mathcal{Q}_p > \mathcal{Q}_{p,crit} (\tau_{orb})$, where the critical tidal $\mathcal{Q}$ factor for the gas giant based on their Eq. 12 is: 

\begin{equation}\label{eqn1}
\mathcal{Q}_{p, crit} (t) = c \, \frac{m_s}{M_p} \left ( \frac{1}{ \tau_{s}} \right ) ^{13/3} t_{dyn} ^{10/3} \, t
\end{equation}

\begin{figure*}\label{fig1}
  \includegraphics[width=\textwidth]{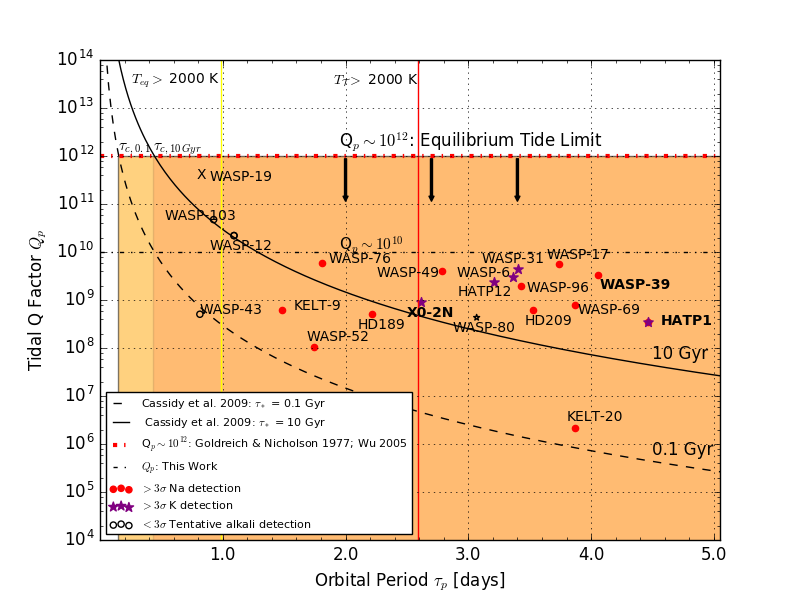}
  \caption{\footnotesize \textbf{Satellite stability diagram of a close-in gas giant exoplanet system.} Critical tidal $\mathcal{Q}$ function, $\mathcal{Q}_{p,crit}$, versus 
  planetary orbital period $\tau_p$ for 100 Myr (dashed) and 10 Gyr (solid) gas giant systems. The equilibrium tidal $\mathcal{Q}_p$ from \citet{goldreich77} and \citet{wu05-1} (horizontal, dash-dot), $\mathcal{Q}_p \sim 10^{12}$, defines the region of stability, $\mathcal{Q}_p > \mathcal{Q}_{p,crit}$, based on the known semimajor axis $a_p$, stellar age t$_*$, and an assumed Io-mass satellite. The tidally-stable region evolves in time towards larger orbital periods, illustrated from light orange ($\tau_{p,crit} (100Myr) \gtrsim 0.2$ days) to dark orange ($\tau_{p,crit} (10Gyr) \gtrsim 0.43$ days). The orange circles and purple stars show the calculated tidal $\mathcal{Q}_{p,crit}$ for all $> 3 \sigma$ detections to our knowledge of Na I \& K I exoplanetary systems respectively. The empty circles and stars, show tentative detections of Na \& K, while X shows that current observations have not detected Na I or K I, yet an atmosphere has been detected. The Na I \& K I observations when examined in conjunction with the tidally-stable region show that all robust detections of Na I \& K I exoplanets can host an Io-mass satellite. The current non-detections (i.e. WASP 19b) could suggest a more moderate stability limit $\mathcal{Q}_p$ $\lesssim 10^{11}$. The tentative detections at ultra-hot Jupiters have high tidal $\mathcal{Q}$s (i.e. WASP-12b, WASP-103b) which could also suggest a mass loss history inside the equilibrium $T_{eq} \gtrsim 2000$K, and tidal $T_{\mathcal{T}} \gtrsim 2000$K (Eqn. \ref{tidalenhancement}) temperature limits computed for a Sun-like star. Within these orbital periods with (red: $\tau_{p} \lesssim 2.6$ days) and without (yellow: $\tau_{p,crit} \lesssim 1$ day)) efficient tidal heating, the extreme mass loss could result in desorbing debris and/or plasma tori. The ultimate survival of exo-Ios is expanded upon in Section \ref{magnetotidal}. }
\end{figure*}

Here m$_s$ and M$_p$ are the satellite and planet masses with dynamical time $t_{dyn} =  2\pi \left ( \frac{R_p^3}{G M_p}\right )^{1/2}$, and c = 1.9 $\times$ 10$^{2}$. Using Eq. \ref{eqn1}, an Io-mass satellite in a stable orbit at 2 R$_p$ around HD189733b gives $\tau_s/\tau_{orb} = 0.18$ hence consistent with the use of $\tau_s/\tau_{orb} = 0.2$ in \citet{cassidy09}. Therefore, accounting for eccentricities driven by the parent star by using Eq. \ref{eqn1}, an Io-like satellite orbiting a Jovian mass exoplanet ($m_s/M_p = 4.7 \times 10^{-5}$), for which $t_{dyn} = 0.12$ days, and $\tau_{orb} = 5\tau_s$ can be sustained beyond a critical orbital period of the planet $\tau_{crit}= 150 \times (t/\mathcal{Q}_{crit,p})^{3/13}$ for t Gyr. This results in a regime stable to orbital decay for 0.1 and 10 Gyr as indicated by the solid and dashed lines in Fig. \ref{fig1}.

\begin{table*}[t]
    \centering
    \hspace*{-1.5cm}
\begin{tabular}{llrrrrrrrrrr}
\hline
 Stellar System   & Type   & T$_*$ &  m$_v$ & $M_{*}$ [M$_{\odot}$] &   M$_{p}$ [M$_J$] &   R$_{*}$ [R$_P$] &   R$_{p}$ [R$_J$] &   $\rho$ [g/cm$^3$] &   t$_*$ [Gyr] &   a$_{p}$ [R$_{*}$] &   $\tau_{p}$ [days] \\
\hline
 WASP-52         & K2V   &5000    & 12.0 &                   0.87 &              0.46 &              6.06 &              1.27 &                0.28 &          0.40 &                7.40 &                1.75 \\
 WASP-76         & F7   &6250     & 9.5 &                   1.46 &              0.92 &              9.20 &              1.83 &                0.19 &          5.00 &                4.10 &                1.81 \\
 HD189733        & K2V  &4875      & 7.66 &                  0.85 &              1.16 &              6.88 &              1.14 &                0.98 &          0.60 &                8.28 &                2.22 \\
 XO-2 N          & K0V  &5340    & 11.18 &                0.98 &              0.62 &              9.64 &              0.97 &                0.83 &          2.00 &                8.23 &                2.62 \\
 WASP-49         & G6V  &5600    & 11.36 &                0.72 &              0.38 &              9.06 &              1.11 &                0.34 &          5.00 &                7.83 &                2.78 \\
 HAT-P-12        & K4   &4650     & 12.84 &               0.73 &              0.21 &              7.14 &              0.95 &                0.30 &          2.50 &               11.80 &                3.21 \\
 WASP-6          & G8V  &5450    & 12.4 &            0.89 &              0.50 &              6.92 &              1.22 &                0.34 &         11.00 &               10.41 &                3.36 \\
 WASP-31         & F    &5540      & 11.7 &                   1.16 &              0.48 &              7.85 &              1.54 &                0.16 &          5.00 &                8.08 &                3.41 \\
 WASP-96         & G8   &5540     & 12.2 &                   1.06 &              0.48 &              8.51 &              1.20 &                0.34 &          8.00 &                9.28 &                3.43 \\
 HD209458        & G0V  &6092    & 7.65 &                   1.13 &              0.69 &              8.48 &              1.38 &                0.33 &          4.00 &                8.04 &                3.50 \\
 WASP-17         & F4   &6650     & 11.6 &                   1.20 &              0.51 &              6.74 &              1.99 &                0.08 &          3.00 &                8.02 &                3.74 \\
 WASP-69         & K5   &4715     & 12.4 &                    0.83 &              0.26 &              7.48 &              1.06 &                0.27 &          2.00 &               11.97 &                3.87 \\
 WASP-39         & G8   &5400     & 12.11 &                    0.93 &              0.28 &              6.86 &              1.27 &                0.17 &          5.00 &               11.68 &                4.06 \\
 HAT P 1        & G0V   &5980    & 10.4 &                    1.15 &              0.53 &              8.66 &              1.32 &                0.29 &          3.60 &               10.19 &                4.47 \\
\hline
\end{tabular}
    \caption{Observed stellar system parameters. Spectral type, V band apparent magnitude m$_v$, stellar mass M$_*$ (in Solar masses: $M_{\odot} = 1.9884 \times 10^{33} g$), planet mass M$_p$ (in Jovian masses M$_J$ =1.8983 $\times 10^{30}$ g),  stellar radius R$_*$ (in planetary radii R$_p$ in Jovian radii R$_J$ = 7.1492 $\times 10^{8} cm $),  planetary density $\rho$ in g cm$^{-3}$, stellar age t$_*$ (if unknown set to average value $\sim 5$ Gyrs), semimajor axis a$_p$ (in stellar radii R$_*$), and observed period $\tau_p$ in days. The tidal $\mathcal{Q}$ calculations rely only on these known quantities.   }
    \label{T2stellar}
\end{table*}

\begin{table*}[t]
    \centering
    \hspace*{-1.5cm}
\begin{tabular}{lrrrrrrrrrr}
\hline
 Planetary System   &   W$_{\lambda,i}$ [m\AA] &   T$_{eq}$ [K] &   H [km] &   $\tau_{i}$ [min] &   v$_{min}$ [km/s] &   R$_{i}$ [R$_p$] &   R$_{Roche}$ [R$_p$] &   a$_{Hill}$ [R$_P$] &   a$_{s}$ [R$_P$] &   $\tau_{s}$ [hours] \\
\hline
 $^{1 \textcolor{red}{\bullet}}$WASP-52b$^{\ominus}$           &                  58 &        1315.00 &   669 &               13.0 (3.5) &               7.82 &               1.07 &                  1.11 &                 2.40 &              1.17 &                 8.40 \\
$^{2 \textcolor{red}{\bullet}}$ \underline{WASP-76b}$^{\ominus}$           &                   2.78  &        2190.00 &  1156 &                1.4 (0.4) &             228.47 &               1.15 &                  0.97 &                 2.21 &              1.08 &                 8.69 \\
$^{3 \textcolor{red}{\bullet}}$ \underline{HD189733b}          &                   6.72  &        1200.00 &   194 &               16.9 (4.5) &              11.40 &               1.14 &                  1.69 &                 4.32 &              2.11 &                10.65 \\
$^{4 \textcolor{red}{\bullet}\textcolor{violet}{\star}}$ XO-2 N b           &                  20 (10) &        1500.00 &   332 &                3.8 (1.02)&               6.81 &               1.02 &                  1.60 &                 4.21 &              2.07 &                12.56 \\
$^{5 \textcolor{red}{\bullet}}$\underline{WASP-49b}$^{\ominus}$           &                  6.05  &        1400.00 &   668 &                4.0 (1.1) &             157.57 &               1.48 &                  1.19 &                 3.56 &              1.74 &                13.35 \\
$^{6 \textcolor{violet}{\star}}$ HAT-P-12b$^{\ominus}$          &                   (160) &         960.00 &   604 &                26.0 (7.0)&               (2.84) &               (1.03) &                  1.14 &                 3.79 &              1.86 &                15.42 \\
$^{7 \textcolor{violet}{\star}}$ WASP-6b$^{\ominus}$            &                   (110) &        1150.00 &   500 &                5.0 (1.3) &                (18.6) &               (1.02) &                  1.19 &                 3.84 &              1.77 &                16.13 \\
$^{8 \textcolor{violet}{\star}}$ WASP-31b$^{\ominus}$           &                   (10) &        1580.00 &  1128 &                2.8 (0.7) &               (59.4) &               (1.44) &                  0.93 &                 3.23 &              1.58 &                16.35 \\
$^{9 \textcolor{red}{\bullet}}$ WASP-96b$^{\ominus}$           &                 110 &        1285.00 &   559 &                5.8 (1.5) &               6.12 &               1.02 &                  1.19 &                 4.29 &              2.10 &                16.44 \\
$^{10 \textcolor{red}{\bullet}}$ \underline{HD209458b}$^{\ominus}$          &                   1.01 &        1450.00 &   580 &                5.7 (1.5) &              13.69 &               1.05 &                  1.17 &                 3.95 &              1.93 &                16.80 \\
$^{11 \textcolor{red}{\bullet}}$ WASP-17b$^{\ominus}$           &                  13 &        1740.00 &  1961 &                3.4 (0.9) &             240.90 &               1.35 &                  0.73 &                 2.70 &              1.28 &                17.93 \\
$^{12 \textcolor{red}{\bullet}}$\underline{WASP-69b}$^{\ominus}$           &                  8.04 &         963.00 &   600 &               35.9 (9.6) &               6.47 &               1.18 &                  1.10 &                 3.97 &              1.94 &                18.57 \\
$^{13 \textcolor{red}{\bullet} \textcolor{violet}{\star}}$ WASP-39b$^{\ominus}$           &                   0.93 (430) &        1120.00 &   936 &                6.7 (1.8) &               6.54 &               1.03 &                  0.94 &                 3.66 &              1.79 &                19.47 \\
$^{14 \textcolor{red}{\bullet} \textcolor{violet}{\star}}$ HAT P 1-b$^{\ominus}$          &                  28 (34) &        1322.00 &   629 &                8.7 (2.3) &              14.95 (12.2) &               1.08 (1.10) &                  1.12 &                 4.65 &              2.28 &                21.46 \\
\hline
\end{tabular}
    \caption{Observed planetary system parameters. W$_{\lambda,i} = \delta_i \Delta \lambda,i$, the equivalent width in m$\AA$ is reported directly from the corresponding Na I ($^{\textcolor{red}{\bullet}}$) (and low-resolution K I ($^{\textcolor{violet}{\star}}$) observations if resolved. If not explicitly stated it is computed based on the spectral resolution $R = \frac{\Delta \lambda}{\lambda}$ and wavelength of Na D2. T$_{eq}$ is the radiative equilibrium temperature for a zero-albedo surface in Kelvins. The corresponding scale height H = $\frac{k_b T_{eq}}{\mu m g }$, is computed for a hydrogen/helium envelope. $\tau_{i}$ alkaline Na I (and K I) lifetime limited by photoionization using rates from \citet{huebnermujherjee} for G stars: $k_{NaD2, G} = 5.92 \times 10^{-6}$ s$^{-1}$. For the F and K stars in our sample we use  $k_{NaD2, F} = 1.3 \times 10^{-5}$ s$^{-1}$ and  $k_{NaD2, K} = 9.5 \times 10^{-7}$ s$^{-1}$ the latter being the value adopted for HD189733b in \citep{huang17}. The minimum, ionization-limited velocity is given by $v_{min} \sim R_{i}/ \tau_{i}$, where $R_{i} =\left( \delta_i R_*^2 + R_p^2  \right )^{1/2}  $ is constrained by the transit depth at line center assuming a spherically symmetric alkaline $i$, envelope as described in the text. For planets with a relatively low R$_{i}$, adequate endogenic explanations exist (e.g.\citet{Sing2016}; \citet{nikolov18})}. The Roche limit, $R_{Roche} = 2.456 R_p \left ( \frac{\rho_p}{\rho_s} \right ) ^{1/3} $, is computed for an Io-like satellite of density $\rho_{Io} = 3.5$ g cm$^{-3}$. The Hill sphere, $a_{Hill} = a_p (1-e_p) \left (\frac{M_p}{3 M_*} \right ) ^{1/3}  $, the minimum satellite semimajor axis $a_s$ (in R$_p$) is computed following \citet{domingos06}, yielding the corresponding minimum satellite orbital period $\tau_s$ in hours. \\
    $^{1}$ \citet{Chen2017} 
    $^{2}$ \citet{Seidel19} 
    $^{3}$ \citet{huitson2012}; \citet{wyttenbach15}; \citet{khalafinejad17}  
    $^{4}$ \citet{sing11}; \citet{sing12_xo2b_na}     
    $^{5}$  \citet{wyttenbach17}
    $^{6}$ \citet{Barstow2017}; \citet{alexoudi18}  
    $^{7}$ \citet{Barstow2017}; \citet{nikolov14}  
    $^{8}$  \citet{sing15_w31b_K}; \citet{gibson19}  
    $^{9}$ \citet{nikolov18} 
    $^{10}$ \citet{charbonneau02}; \citet{Snellen2008}; \citet{Langland-Shula09}; \citet{Vidal-Madjar2011}; \citet{Sing2016}  
    $^{11}$ \citet{Barstow2017}; \citet{Khalafinejad2018}  
    $^{12}$ \citet{Casasayas-Barris2017} 
    $^{13}$ \citet{Nikolov2016}; \citet{Fischer2016} 
    $^{14}$  \citet{Wilson2015}; \citet{Sing2016} 
    \label{T3planet}
    %
\end{table*}

Using the large equilibrium tide value, $\mathcal{Q}_{crit,p} =10^{12}$ \citep{goldreich77}, $\tau_{crit, 10 Gyr} \gtrsim 0.43$ days, or $\tau_{crit, 0.1 Gyr} \gtrsim 0.2$ days, much closer-in than our sample of hot Jupiters, as indicated by the shaded regions in Figure \ref{fig1}. All of the observed Na I (red points) and K I (purple points) exoplanets reside in a region where an Io-mass exomoon can be dynamically stable throughout the lifetime of the stellar system unless $\mathcal{Q}_p$ exceeds $\mathcal{Q}_{p,crit}$ as given above. Determining this region of stability could also be suggestive of the unknown tidal $\mathcal{Q}$. Our calculations in Figure \ref{fig1} suggest a more moderate range than the equilibrium tide limit of $\sim 10^{12}$, rather a $\mathcal{Q}_{crit, J-exoIo} \sim 10^6 - 10 ^{10}$ for a hot Jupiter-exo-Io system. The lower tidal $\mathcal{Q}$ may explain the lack of robust Na detections for planets outside the inner 'stable exo-Ios' regime. Since turbulent viscosity increases with heat flux, and the large radii of many hot Jupiters might indicate larger internal heat flux, a more moderate $\mathcal{Q}_{crit}$, could be expected.

\indent Given that a stable exomoon uniquely orbits at close proximity \citep{domingos06}, survival against gas drag of an extended and ionized atmosphere should be considered (see Fig. \ref{exoarch}). At $\sim$ 2$R_J$ the plasma density is $\lesssim 10^7$ cm$^{-3}$ \citep{huang17}, where the relative plasma-satellite velocity is $\sim 10$ km/s. Using an expression for the drag force with a drag coefficient that equals one (\citet{passy12}) the time for orbital decay is of the order of, $\tau_{drag} \sim \frac{L}{T}$. Here $T$ is the gas torque from the 
ionized escaping atmosphere on a satellite with angular momentum $L$, giving a $\tau_{drag} \gtrsim 200 Gyr$ which is a few orders of magnitude longer than the lifetime of the stellar system. Thus its contribution to orbital decay is negligible. Moreover we note that while we focus on Io throughout this work, the expected thermal and plasma-driven mass loss described in Section \ref{magnetotidal}, imply the radius of an Io could erode to an Enceladus-mass satellite affecting the $\mathcal{Q}_{p, crit}$ needed on the order of magnitude level, thereby extending the region of stability to closer-in orbits (Fig.\ref{fig1}). Given that all alkaline exoplanets can host an Io-mass satellite, in principle, the absorption could be evidence of tidal activity from a satellite. Markedly, a benchmark case of Na I absorption in transmission spectroscopy has been recently observed in Jupiter's upper atmosphere presenting the opportunity to evaluate the endogenic and/or exogenic origin of Na I. In the following, we therefore consider the sources of Na I required to supply the observed line-of-sight column density derived from the Na I flux decrease in transit as a guide. 
\section{Exogenic Sodium in Jupiter's Atmosphere}\label{jupitersNa} 
Atomic lines of Na I (and several other species, such as Mg I, Fe I, Ca I, Mn I, Li I, Cr I) were briefly ($<$ 1 hour) detected in emission from the plumes associated with the impacts of the largest fragments of comet Shoemaker-Levy 9 (SL9) at Jupiter in July 1994 (see review in \citet{crovisier96}). Most of these emission lines are also observed in the spectra of sungrazing comets, shedding little doubt that the atoms responsible for these atomic lines were present in the impactor itself. Although masses of deposited elements have been quoted in some papers (e.g. \citet{noll95}), they are quite uncertain because (i) they assume non-saturated lines, where optical depth effects are difficult to assess and could lead to underestimating the abundances by several orders of magnitude (ii) they were derived upon the assumption that resonant fluorescence is the only excitation mechanism; in reality, other mechanisms, such as thermal excitation by collisions, electronic recombination and prompt emissions may also contribute \citep{crovisier96}.

\begin{figure*}\label{fig3}
  \includegraphics[width=\textwidth, trim=0cm 14cm 0cm 7cm]{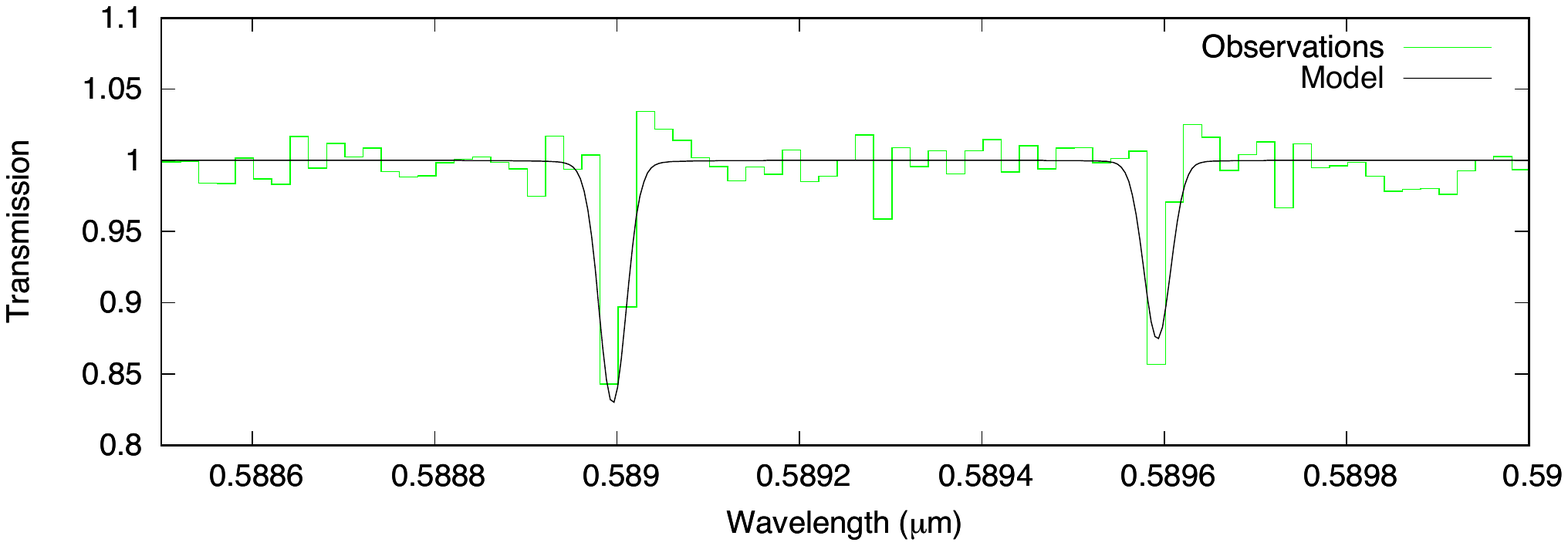}
  \caption{Jupiter's calculated transmission spectrum in the vicinity of the Na D lines for tangent level at 30 mbar (black line). The line-of-sight Na column density is 7$\times$10$^{11}$ cm$^{-2}$ and the spectral resolution is 17500. The model provides a good match of the observed spectrum
(green line) by \citet{montanes} when Ganymede was in eclipse (their Fig. 4).   }
\end{figure*}

Recently, \citet{montanes} (their Fig.4) obtained the first transmission spectrum of Jupiter's limb by observing an eclipse of Ganymede as it entered Jupiter's shadow. The observation was carried out with VLT/XSHOOTER at a spectral resolution of $\sim 0.34 \AA$. Jupiter's cold atmosphere at T$\sim 170 K$ absorbed $\sim 15$ \% of the planet's continuum flux at the Na D1 and D2 doublet during transit. This first detection of Na I in Jupiter's upper atmosphere offers an avenue for endogenic-exogenic comparisons to hot Jupiters. We perform a simple model based on the contribution functions (Fig. 3 \citet{montanes}) at P$_{tot}$ = 30 mbar. For the Na lines, we used spectroscopic parameters kindly provided to us by B. B\'ezard, including a collisional half-width of 0.270 cm$^{-1}$/atm at 300 K. For simplicity, the entire stratosphere above 30 mbar was assumed isothermal at 170 K. Fig. \ref{fig3} shows our modeled spectrum for a line-of-sight (LOS) Na column density of 7$\times$10$^{11}$ cm$^{-2}$ fortuitously close to the optically-thin column density based on the equivalent width W$_{\lambda,i}$ of the spectral line (Eqn \ref{Eqn:obscolumn}). Accounting for the Chapman enhancement factor $\sqrt{2\pi \frac{R_J}{H}}$  $\sim$ 132, where $R_J$ is Jupiter's radius and $H \sim 25$km the atmospheric scale height in the region of interest, gives a radial column density of $\sim$5$\times$10$^9$ Na cm$^{-2}$ above 30 mbar, i.e. 
a volume mixing ratio\footnote{ N.B: Solar abundance: $\left ( Na/H_2) \right)_{\odot} \sim 10^{-6}$ or $X_{Na, \odot} \sim$ 1.7 ppm \citealt{asplund2009} is assumed in endogenic Na models } Na/H$_2$ $\sim$ 10$^{-15}$ or $X_{Na} \sim 10^{-9}$ ppm (demonstrating the extraordinary sensitivity of the lunar eclipse technique by \citet{palle09}; \citet{montanes}). 
We assume below that this mixing ratio holds to deeper levels, as a result of vertical mixing. \\

We note that a Na/H$_2$ volumetric mixing ratio of 10$^{-15}$ at the 30 mbar level corresponds to a partial pressure of Na of $\sim 3 \times 10^{-17}$ bar. Although the vapor pressure of Na is not well measured at low temperatures, this most likely implies a strong supersaturation of Na. Using the recommended Na vapor pressure expression $P_{sat}(bar) = 10^{(5.298 - 5603/T(K))}$ by the CRC handbook of Chemistry and Physics \citep{lide94}, yields $\sim 2 \times 10^{-28}$ bar at 170 K, implying that the derived partial pressure of $\sim 3 \times$ 10$^{-17}$ bar, requires equilibrium at $\sim$ 255 K. Such a temperature is reached in Jupiter's atmosphere near the 0.25 $\mu$bar region. If the Na detected by \citet{montanes} lay in the sub-microbar region, the required Na LOS column density would be considerably higher than $7 \times 10^{11}$ cm$^{-2}$ because at these lower pressures, pressure broadening becomes unimportant, leading to more saturated and narrower lines requiring higher opacities. It is striking however that the strongest Na lines observed by \citet{montanes} occur during the umbra (their Fig. 4). This suggests that the bulk of the Na lies in the atmospheric region sampled during the eclipse, i.e. near $\sim 30$ mbar, and a possibility would be that Na is supersaturated in Jupiter's stratosphere, perhaps due to inefficient condensation on aerosols.

Gaseous Na observed in Jupiter's stratosphere is most certainly of external origin given that any Na-bearing species originating from the interior are expected to condense out in the cold upper troposphere. To assess the external flux required by the observations, we followed the simple approach of \citet{bezard2002}, in which the deposition rate of Na is balanced by its removal from the stratosphere by vertical transport.
We used an eddy diffusion coefficient $K$ increasing as n$^{-1/2}$ (where n is the atmospheric number density),
above a pressure level $P_0$ = 300 mbar, where $K$ has a minimum value K$_0$ = 700 cm$^2$s$^{-1}$. Following \citet{bezard2002} (Eqn.3) where the column density of CO above P$_0$ was derived based on the source flux $\Phi$, we rewrite this expression based on a spherically-symmetric mass accretion rate of an exogenic source $\dot{M}_{exo}$ yielding a radial column density: 

\begin{equation}\label{Eqn:exogenicaccretion}
N_{exo}= \frac{\dot{M}_{exo} H^2}{2 \pi R_p^2 m_i K_0}
\end{equation}

which corresponds to $\sim 5 \times 10^{10}$ Na cm$^{-2}$ above P$_0$ = 300 mbar. Solving for the supply rate we obtain $\dot{M}_{exo} \sim$ 0.05 g/s. This meager mass supply can be understood in terms of the total volatile mass in the envelope. Using the observed column density at 30 mbars: $N_{exo} \sim$5$\times$10$^9$ cm$^{-2}$ we find a total mass of $\approx$ 1.2$\times$10$^{8}$g of Na integrated over Jupiter. Despite the simplicity of the model, this number is likely accurate to within a factor of a few.

\subsection{Exogenic Sources}\label{exogenicsources}
Possible external sources for the Na I currently observed in Jupiter include: (i) the Shoemaker-Levy 9 (SL9) collision (ii) past cometary impacts (iii) micrometeoritic flux and (iv) Na I escaping from Io's atmosphere. We demonstrate below that each of these sources is in fact sufficient to account for the above accretion rate $ \dot{M}_{exo}$.

\subsubsection{Shoemaker-Levy 9 Impacts}
The best determined amount of materials delivered/produced at Jupiter by the SL9 impacts is for the long-lived species (CO, HCN, CS), as
those were observed both just after the impacts (see review \citet{lellouch96}) as well as monitored on yearly timescales. The mass of CO produced in particular, has been estimated to be typically $\sim$5$\times$10$^{14}$g (\citet{lellouch97}; \citet{moreno03}), with a probable uncertainty better than a factor-of-two. This figure corresponds to $\sim$2.2$\times$10$^{14}$g of atomic O, of cometary origin. The Na/O ratio for cosmic abundance can be taken as Na / O = 3.6$\times$10$^{-3}$ (recommended value from \citet{lodders2010}) \footnote{Lodders indicates Na/O = 7.5$\times$10$^{-3}$
for chondritic composition and Na/O = 3.7$\times$10$^{-3}$ in the Sun}. Assuming cosmic abundances in SL9 would thus imply a mass of $\sim$8$\times$10$^{11}$g of Na deposited by the impacts. Given our estimate of $M_{Na}$ above, it would appear that impacts can easily source the required Na. However, the SL9-delivered material was primarily deposited at the $\sim$0.1 mbar level (e.g. \citet{lellouch97}) and that as of today, it has likely diffused down to pressures of only a few millibars. Thus a constant Na source today from the SL9 impacts is unlikely as the Na is observed to be present down to at least 30 mbar.

\subsubsection{Older Cometary Impacts}
\citet{bezard2002} showed that CO in Jupiter actually originates from three different components (i) internal CO (ii) CO deposited by the SL9 impacts (ii) additional stratospheric CO, most probably due to the deposition by a suite of "old" cometary impacts with 
some size distribution for the impactors, which has by now invaded the entire stratosphere. The third component is associated with
$\sim$ 44 - 300 kg/s of CO. Rescaling the above by the Na/O cosmic ratio, suggests that these old impacts
would have additionally delivered $\sim 0.13-0.9$ kg/s more than sufficient to explain the equivalent mass of Na observed.

\subsubsection{Cosmic Dust} 
Micrometeoritic flux is also a permanent source of external material, especially oxygen, to the outer planets (see \citet{mosespoppe2017} and references therein). These interplanetary dust particles (IDP) originate from comet activity or disruption, from collisions in the asteroid and Kuiper belts, and from interstellar dust particles streaming into the solar system. Even though comets are an important source of micrometeorites, an important difference for planetary atmospheres is that micrometeoritic impacts are thought to preserve the H$_2$O molecules upon entry and ablation, while cometary impacts convert cometary H$_2$O into atmospheric CO due to shock chemistry. At Jupiter, based on the observed H$_2$O vertical profile in 1997, which was essentially consistent with expectations from a SL9 source, \citet{lellouch02} found that the production of H$_2$O due to permanent micrometeoritic influx is less than $\sim$1.5 kg/s , with a best-guess estimated value of $\sim$0.75 kg/s. This surprisingly low value, compared to expectations from the dynamical dust model of \citet{poppe2016} (who predict typically 2 orders of magnitude larger fluxes) may suggest that a dominant fraction of the H$_2$O contained in IDPs is in fact {\em not} preserved at micrometeoritic entry. Notwithstanding with this poorly understood issue (see also \citet{mosespoppe2017}), we note that even a scaling of the above H$_2$O production rate from micrometeorites by the Na/O cosmic abundance yields a Na deposition rate of $\sim$ 3.7 g/s, once again more than sufficient to explain the required Na production rate of $\sim$ 0.05 g/s. 
\subsubsection{Sodium Accretion from Io}
Finally, we calculate the Na I deposition rate on Jupiter from Io, the focus of our work. The accretion rate onto Jupiter assuming the Na I is escaping isotropically by the physical processes we describe in Section \ref{magnetotidal} is: $\dot{M}_{acc} \sim \left ( \frac{R_J}{a_s} \right )^2\dot{M}_{Io} $. For a lower limit to the Na mass loss rate at Io, $\dot{M}_{Io} \gtrsim$ 10 kg/s and a semimajor axis of $a_s$ = 5.9 R$_J$ the Na accretion rate is $0.1$ kg/s, roughly 2000 times larger than required by the observations. In the anisotropic case, the Na could be collimated into a jet leaving Io at $\sim$ 3.8 g/s \citep{schmidt15AGU}, which would accrete onto Jupiter at a more moderate rate of $\sim 0.11$ g/s, yielding the observed mass flux within a factor of 2. We illustrate the physical process of such jets in Figure \ref{Nanebula}. As discussed below, this source is evidently variable, but appears to be quite consistent with the observations. All exogenic sources are capable of supplying considerable amounts of atomic Na I, possibly in excess of the minimum abundance. This may suggest a dominant fraction of the incoming Na flux is chemically converted to other Na-bearing species, with NaH being a logical candidate. \\

Besides Na I, recent observations by Juno suggest that Io's jets are spraying Jupiter with volatiles based on the detection of sulfur and oxygen \textit{ions} (S III, O II) in Jupiter's atmosphere at equatorial latitudes \citep{valek18} strongly complementing the exogenic nature of Na we propose. The extreme tidal heating and mass loss at exo-Ios we will now describe may maintain a viable source of Na to a hot Jupiter's upper atmosphere.

\section{Exogenic Sodium in Jupiter's Exosphere: Satellite Atmospheric Escape}\label{magnetotidal}
Io's geologic activity, which results in the ejection of NaCl \& KCl into Io's atmosphere is driven by tidally-heated volcanism whereas the observed and widely distributed Na I \& K I are primarily produced by the interaction of the volatiles with the plasma flow in Jupiter's $\sim \,$ 4.17 Gauss magnetic field.  The net plasma flow, which governs the momentum and eventual knock-off of volatiles from the moon as we will describe, is set by the velocity difference between the plasma torus and the outgassing moon. This feedback process has produced a number of observed alkaline features driven by several molecular physics processes (e.g., \citet{wilson02}). Although this has resulted in an extensive literature, our focus is on determining the ability of a close-in exoplanetary system to invigorate a volcanic moon and drive a range of possible alkaline source rates to the system by scaling to our robust understanding of the Jupiter-Io system. The resulting column densities can then be directly compared to the observed equivalent widths of alkalis at exoplanets, W$_{\lambda,i}$ (Table \ref{T3planet}). We describe the mass loss of Na I \& K I at an Exo-Io, generalized for a close-in rocky body subject to tidal heating and irradiation. As potassium's signature as a volcanic alkali is similar at Io \citep{thomas96}, we focus the following in terms of the robust spectral observations of sodium for simplicity. \\

\subsection{Tidally-Driven Volcanism at an Exo-Io}\label{tidally-driven}
Tidal heating of Io has been shown to be responsible for its widespread volcanism. The tidal heating rate of Jupiter's tidally-locked moon, $\dot{E}_s \propto \frac{e^2}{Q_s}$ driven by forced eccentricities, $e$, locked by Europa and Ganymede's Laplace Resonance with Io, is the dominant interior heating source. Similarly, the tidal heating of an exomoon will likely dominate the interior energy budget due to the additional \textit{stellar} tide. Consequently the tidal heating rate is orders of magnitude higher than at Io, which for an exo-Io of similar rheological properties ($\mathcal{Q}_s \sim 100$, $R_s = R_{Io}$, $\rho_s = \rho_{Io}$), can be written as following \citep{cassidy09} (Eqn. 19\& 20) as: 

\begin{equation}\label{tidalheating}
\dot{E}_s = \frac{\upsilon}{\mathcal{Q}_s} \frac{\rho_s ^2 R_s ^7}{\tau_p ^4 \tau_s}
\end{equation}

where $\upsilon = 3 \times 10^{-7}$ cm$^{3}$ erg$^{-1}$, and $\tau_s= \tau_p/5$ based on the tidal stability criterion discussed in Section \ref{tidalstability}. For utility, we describe the exo-Io's tidal efficiency as: $\eta_{\mathcal{T}} = \frac{\dot{E_s}}{\dot{E_{Io}}}$, which can readily be computed for any 3-body system as tabulated in Table \ref{T4sodiumpotassium}. The enhanced tidal-heating described in Eqn. \ref{tidalheating} will also contribute to a tidally-heated surface temperature $T_0 = T_{eq} + \Delta T_0$ which is very roughly approximated as: 

\begin{equation}\label{tidalenhancement}
\Delta T_0 = \left ( \frac{\dot{E_s}}{4 \pi R_s^2 \sigma_{sb}} \right ) ^{1/4}
\end{equation}

where $\sigma_{sb}$ is the Stefan-Boltzmann constant. 
At Io, the total neutral volcanic content (SO$_2$, SO, NaCl, KCl, Cl, and dissociation products) ejected to space (Section \ref{plasmadriven}) by the incident plasma is estimated to be on average $\sim 1000$ kg/s (e.g. \citet{thomas04}) varying within an order of magnitude over decades of observations (\citet{burger01}; \citet{wilson02}; \citet{thomas04}). While the source of the dominant gas SO$_2$ is ultimately tidally-driven volcanism, the near-surface atmosphere is mostly dominated by the sublimation of SO$_2$ frost \citep{tsang16}. 
By observing the atmospheric evolution of the SO$_2$ column density with heliocentric distance, \citet{tsang13} estimated the direct volcanic component to be $N_{volc} \sim 6.5 \times$ 10$^{16}$ cm$^{-2}$ typically $\sim \frac{1}{3}$ of the total observed SO$_2$ column density. \citet{ingersoll89} demonstrated the relative contributions due to both sublimation and volcanic sources in maintaining Io's atmosphere, and established a relationship relating the volcanic source rate to the volcanically-supplied atmospheric pressure:
\begin{equation}
P_{volc} = \frac{v}{\sqrt{32} \pi \alpha R_{Io} ^2  } \dot{M}_{0,volc: Io} 
\end{equation}
This expression is also equivalent to the volcanic column density $N_{volc} = \frac{P_{volc}}{m_i g}$, where g is the acceleration due to gravity. Adopting an observed atmospheric temperature of T$_{atm} = 170$ K by \citet{lellouch15} corresponding to an atmospheric scale height of $H = 12$km,  thermal velocity $v = \sqrt{gH}$ equal to 150 m/s, and a sticking coefficient $\alpha = 0.5$ for the SO$_2$ mass of 64 amu yields a volcanic source rate of $\dot{M}_{0,volc: Io} \sim 6.9 \times 10^6$ kg/s of SO$_2$ integrated over Io's mass M$_{Io}$. The average volumetric mixing ratio for NaCl to SO$_2$ at Io is observed to be $X_{NaCl} \sim 3 \times$ 10$^{-3}$ \citep{lellouch03}. This leads to a source rate of $\dot{M}_{0,volc: Io} \sim 7.4 \times$ 10$^{3}$ kg/s of NaCl, somewhat larger, yet reasonably consistent with the direct measurement of the NaCl volcanic source rate of 0.8-3.1 $\times$ 10$^{3}$ kg/s \citep{lellouch03}. From these estimates, we will adopt $\sim 3 \times 10^3$ kg/s of Na I as the volcanic source rate for Io. 

This source rate is \textit{tidally-driven}. If this source rate was tidally-limited with an efficiency of 1 (as described further in Section \ref{destruction}), the implied heating rate is within a factor of $\sim 5$ of Io's theoretically-derived $\dot{E} = 1.6 \times 10^{19}$ (\citet{peale79}) validating the ability of $\eta_{\mathcal{T}}$ to probe a supply rate efficiency. 
\subsection{Exo-Io Mass Loss Processes: Exogenic Sodium and Potassium Sources in an Exoplanet Magnetosphere }\label{systemexogenic}

\begin{figure*}\label{Nanebula}
  \includegraphics[width=\textwidth]{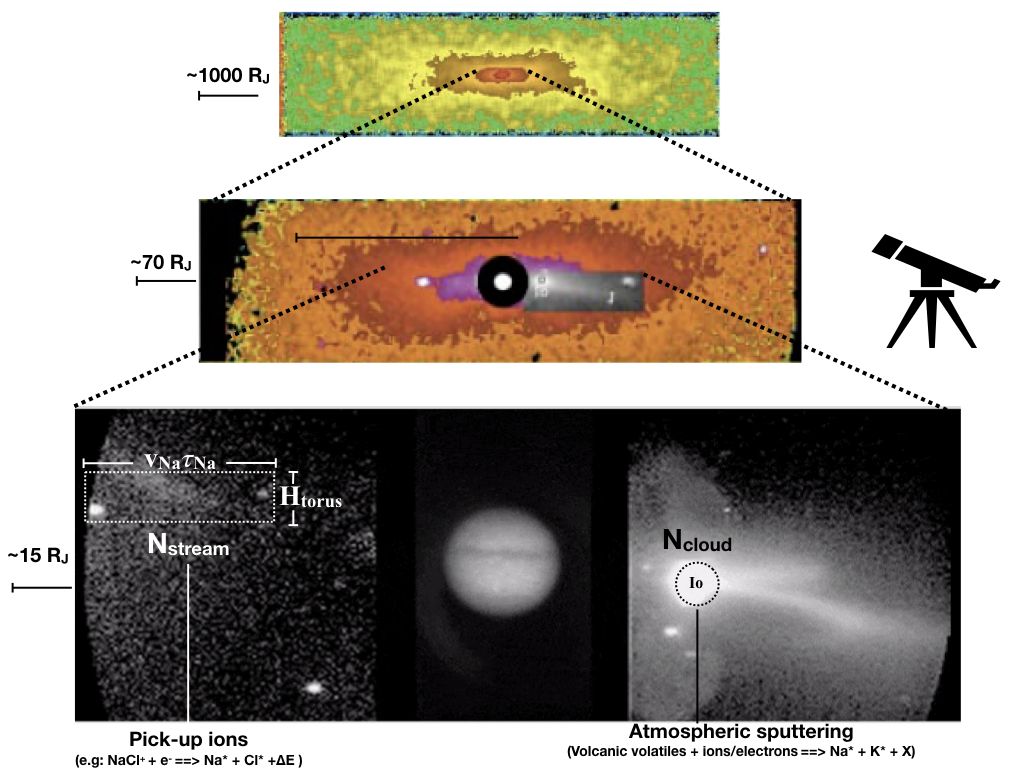}
   \caption{\textbf{2-D ``Face-On'' Architecture of a Sodium Exosphere \textit{imaged} at the Jupiter System. } The rectangular Na I exosphere beyond Io at $15 R_J$,  $70 R_J$, and $1000 R_J$ is adopted from a 1995 emission observation (Fig.8 of  \citet{wilson02}). In the image sequence at Io's orbit observed by \citet{schneider91} fast Na streams $< 100$ km/s are seen to emit from the plasma torus (the data can be accessed on Apurva Oza`s webpage under ``exo-Io"). We identify the two principal Na I features at Io's orbit capable of contributing to a close-in gas giant exosphere: a local cloud and a plasma-driven stream. N$_{cloud}$ (black circle) is driven by atmospheric sputtering at $\sim 10 km/s$, where the radial profile corresponds to \citet{burger01} (similar to Eqn. \ref{eqnexoio}). N$_{stream}$ can be derived based on a cross-sectional stream (white rectangle) extended to a length scale $\sim v\tau_{Na}$, with height equivalent to that of the ion torus $\sim a_s \frac{v_+}{v_{orb}}$, driving the stream. As described in the text, if the entirety of the Jupiter system were to transit the Sun the total LOS column density would be $\sim 2.5 \times 10^{10}$ Na cm$^{-2}$ easily discernable by current ground-based transmission spectroscopy.    }

\end{figure*}

Moons are well known to be significant plasma sources in gas giant magnetospheres (\citet{johnson06a}; \citet{johnson06b}). At Io the ejected species are eventually ionized by the plasma and radiation environment forming a toroidal plasma co-rotating with the Jovian magnetic field which has been remarkably imaged in a sequence of S II \& O II spectra gyrating with Jupiter's B-field (see Fig.1 \citet{schneider91}) and its corresponding Na features which we identify in the context of an exo-Io transit in Fig. \ref{Nanebula}. The plasma interaction with Io's upper atmosphere produces a number of distinguishable Na features produced by a multitude of physical processes including ion-neutral momentum transfer, charge exchange and dissociative recombination of NaCl$^{+}$ as analyzed and simulated by \citet{wilson02}. In Fig. \ref{Nanebula} we identify three principal Na features in Jupiter's exosphere, all ultimately Iogenic \footnote{sourced by Io}: cloud, stream, and nebula. At an exoplanet these features are averaged over the planet's transit duration and are spatially unresolved. Therefore, an estimate of the line-of-sight (LOS) column density \textit{independent} of the spatial distribution is needed during an exoplanet transit. The average LOS column density $ \langle N \rangle $ is simply the total number of atoms $\mathcal{N}$ in the system which is obtained by integrating the LOS column over the surface area of the star:

\begin{equation}\label{eqn:Ncolumn}
\langle N \rangle = \frac{\mathcal{N}}{\pi R_{*}^2}   
\end{equation}

As a thought experiment, the average Na I column density an external observer would notice if Jupiter's exosphere ($\sim 1000 R_J$) were transiting the Sun is $\sim 2.5 \times 10^{10}$ Na cm$^{-2}$. This quantity using the inset image from Fig. \ref{Nanebula} is within a factor of 3 of Jupiter's atmospheric sodium (Section \ref{jupitersNa}).  \\
At a close-in exoplanet system the total quantity of Na atoms can be estimated based on a model of tidally-driven mass loss and ionization as we will describe in Section \ref{columnsection}. Due to ionization and tidal heating the close stellar proximity will simultaneously compactify the overall Na exosphere by $\propto \frac{1}{a_p^2}$ and magnify the overall Na source rate by $\propto a_p ^{15/2}$. Furthermore, as we will describe the minimum observed quantity of Na atoms can also be estimated if the equivalent width of the spectral line is resolved in high-resolution transmission spectroscopy. 

Due to the tilt of the magnetic equator to Io's orbit and the variation in the sourcing, these features are also variable. Only recently has a magnetic field been observed around substellar bodies; L and T dwarfs \citep{kao18}. While an exoplanetary magnetic field is expected, it is still unconstrained (\citet{christensen09}; \citet{matsakos15}; \citet{kislyakova14} ; \citet{Griessmeier07} ; \citet{Rogers17} ). The activity from the type of satellite we describe here could be used to indicate the presence of such a field. It is clear that a description of the magnetospheric environment is needed before assessing the nature of the Na I and K I absorption features seen at hot Jupiters. Due to the shorter photoionization lifetime of the alkalis at a close-in exoplanet ($\tau_{i}; $Table \ref{T3planet} scaled to \citet{huebnermujherjee}), the Na I features seen at an exo-Io would be far smaller than the imaged Jovian system (Figure \ref{Nanebula}). For example, assuming a spherically symmetric endogenic Na I cloud of apparent radius $R_{Na} \sim 1.14 R_p $ (c.f. Table \ref{T3planet}) about the exoplanet HD189733b corresponds to an absorbing layer of area $\sim 6 \times 10^{19}$ cm $^2$ compared to Jupiter's magnanimous Na I cloud $\sim 10^{24}$ cm $^2$ evaluated at radius 70 $R_J$. If Jupiter were close-in, the photoionization time will significantly reduce this area to $\pi R_c^2 \sim \pi (v_{i} \tau_{i})^2 \sim 3 \times 10^{18}$ cm $^2$ when considering a Na I cloud outgassed at $v_{i} \sim 10 km/s$ due to an exo-Io \citep{johnsonhuggins06}. In the following guided by molecular kinetic simulations and Io observations described above, we estimate an exo-Io's Na I escape rate based on its volcanic gas source rate. This will provide rough lower limits on possible alkaline column densities based on our knowledge of tidal heating and thermal desorption from a close-in rocky body.

We will rely on molecular kinetic simulations and Io's observations described above to estimate an exo-Io's Na I escape rate based on its volcanic gas source rate to provide rough lower limits on possible alkaline column densities based on our knowledge of tidal and thermal heating, relevant for a close-in rocky body. In all scenarios we self-consistently include the expected tidal heating based on the above discussion.We focus on three principal drivers of atmospheric escape described in \citet{johnson15}; \citet{johnson04}. 


\subsubsection{\textbf{$\dot{M_P}$} Plasma-driven escape: Atmospheric Sputtering of an Exo-Io}\label{plasmadriven}

Io's volcanic mass loss is due to momentum exchange with the incident plasma, and is often referred to as \textit{atmospheric sputtering} \citep{haff81}. Here, sputtering is used as a proxy to account for net loss due to the ambient plasma and its accompanying fields. Scaling to the plasma pressure at Io, $P_{Io} \sim 1.8 \times 10^{-5}$ dynes cm$^{-2}$ \citet{bob90book} one can derive the tidally-driven atmospheric sputtering of atomic Na I at an exomoon building on \citet{johnson04} as,  

\begin{equation}\label{dmdtP}
\dot{M}_{P} \sim x_i \frac{P_{s}}{P_{Io}} \frac{U(R_x)_{Io}}{U(R_x)_s} \frac{R_{x,s}^2}{R_{x,Io}^2} \dot{M}_{Io}
\end{equation}
which is scaled to Io's measured mass loss rate, $\dot{M}_{Io} \sim 1000 kg/s$. This is generalized for a volatile species $i$, where $x_i$ is the mass fraction. This is set to $0.1$ in the exobase region using the 1995 Na I emission (Figure \ref{exoarch}) based on the total SO$_2$ mass, $P_{s}$ is the plasma pressure at the extrasolar satellite, $U(R_x)$ the satellite's gravitational binding energy at its exobase of radius, R$_x,s$. A lower bound to the exobase based on an exponential atmosphere with a scale height H$_0$ determined by the surface temperature and a surface density, n$_0$, that is enhanced by the extreme tidal heating is:

\begin{equation}\label{Eqn:Rx}
R_{x, s} \sim  R_s + H_0 \, ln \left (n_0  H_0  \sigma_i \right)
\end{equation}

Here, R$_s$, is the satellite radius, H$_0$ the volcanic (SO$_2$) scale height at the tidally-heated surface temperature T$_0$, and $\sigma_i$ the collisional cross section\footnote{for the dominant volcanic species we use: $\sigma_{SO2} = 1.62\times 10^{-14}$ cm $^{2}$} between the escaping alkaline species and ambient atmosphere. Table \ref{T4sodiumpotassium} tabulates lower and upper bounds to the exobase, for an exponential atmosphere.

Since the loss due to ionization occurs in the exobase region, in Eqn. \ref{Eqn:Rx} we use for simplicity: \\  $n_0 \sim \frac{\tau_{Na, exo-Io}} {\tau_{Na, Io}} x_{Na} \eta_{\mathcal{T}} n_{0, SO2}$, where $\frac{\tau_{Na, exo-Io}}{\tau_{Na, Io}}$, accounts for rapid ionization at an exo-Io based on Io's Na lifetime of $\tau_{Na, Io} \sim 4$ hours, and $\eta_{\mathcal{T}}$ is a factor accounting for the enhanced tidal-heating (Equation \ref{tidalheating}). This heating would enhance the surface column density at Io, which we approximate here as the product of the surface number density n$_{0, SO2} = 1.8 \times 10^{11}$SO$_2$ cm$^{-3}$ and the scale height \citep{lellouch15}. We use a near-surface mixing ratio of X$_{Na} = 0.013$ from the direct detection of NaCl accounting for venting \citep{lellouch03}. 

Io's exobase is $\sim 465$ km or $\approx 1.25$R$_{Io}$ (\citet{mcdoniel17}; \citet{wongjohnson96a}; \citet{wongjohnson96b}) whereas the observed Na I profile is far more extended and not exponential (c.f. Section \ref{discussion} Eqn. \ref{exoiow49} and \citet{burger01}). Such an atmospheric tail is not unlike that for a comet or disintegrating rocky body as we describe in Section \ref{destruction}.
However, the exobase altitude is limited by its Hill sphere, so that $R_{x, s} \sim a_{Hill, s}$ in which case the satellite atmosphere experiences Roche-lobe overflow. The satellite Hill radius $a_{Hill, s} \sim a_s \left ( \frac{M_{Io}}{3 M_p}\right ) ^{1/3}$ is tabulated in \ref{T4sodiumpotassium} where the exo-Io is assumed to be in a circular orbit of semimamjor axis a$_s$ (c.f. Table \ref{T2stellar}).


The plasma pressure at the exobase can be roughly written as $P_{p} = P_{mag} + P_{thermal} + P_{ram}$ where the magnetic and ram pressures dominate at Io. Therefore, a lower bound at an exo-Io is obtained using $P_s \sim P_{mag} + P_{ram}  \sim \frac{B_r^2}{2 \mu_0} + n_i m_i (u_i)^2 $, where B$_r$ is the planetary magnetic field strength at the orbit of the satellite, $\mu_0$ the permeability of the vacuum,   n$_i$, m$_i$ and u$_i$ are the ion number density, average mass, and ion flow speed respectively. At an exo-Io the pressure could be larger based on the unknown magnetic field strength of the gas giant. As a lower limit, we assume an unmagnetized gas giant where in Table \ref{T4sodiumpotassium} we calculate the ram pressure at each exoplanet, based on Parker's solar wind model \citep{parker64}, scaling to the ion density at 1 AU. We find that in the unlikely absence of a magnetic field on the gas giant, a satellite would experience a stellar wind ram pressure comparable to the ram pressure at Io’s exobase as indicated by the parameter f$_{ram}$ (Table \ref{T4sodiumpotassium}). As an upper limit we assume a Jovian-like B-field leading to a large magnetic pressure as is the case at Io. At an exo-Io $\sim 2 R_p$ away it is $\sim 700$ that at Io as indicated by the parameter f$_{mag}$ (Table \ref{T4sodiumpotassium}) . We find plasma-driven loss dominates at most alkaline systems with the exceptions of three WASP systems, 52-b, 76-b, and 69-b (energy limited escape) and HAT P 12-b (thermal evaporation), driven by the mechanisms which we shall now describe.


\subsubsection{\textbf{$\dot{M_U}$}: Energy-limited escape from an Exo-Io}
We evaluate the simple energy-limited escape \citep{watson81} regime shown to dominate close-in exoplanet atmospheric escape in the Kepler data (\citet{jinmordasini14}; \citet{fulton18}). Due to the extreme irradiation at a close-in gas giant, the incoming UV and X-ray (XUV) radiation will heat the upper atmosphere of the outgassing exo-Io. The heating will expand the gas and result in hydrodynamic escape which is typically approximated as:

\begin{equation}\label{dmdtU}
\dot{M}_{U} \sim m_i x_i \frac{Q}{U_s (R_{a})}
\end{equation}

where $Q = \eta_{XUV} 4 \pi R_a ^2 F_{XUV}$ is the heating rate due to the incident XUV flux $F_{XUV}$, R$_a$, m$_i$ the volatile mass, x$_i$ the mass fraction, and U$_s$ the binding energy as described in \citet{johnson15} for small bodies. Absorption in the upper atmosphere should be efficient due to the presence of volcanic molecules. In \citet{jinmordasini18} a heating efficiency of 0.1 is used for H/He atmospheres, and for Io's volcanically-generated atmosphere \citet{lellouch92} used a heating efficiency of 0.35. To account for both the expansion of the atmosphere and the heating efficiency, for simplicity we use R$_a \sim$ R$_s$ and $\eta_{XUV} = 0.1-0.4$ (c.f. \citet{mc09}). We caution that the escape may be further reduced due to transonic escape, which occurs when $Q > Q_c$ (c.f. Eqn. 10; and Fig. 2 \citet{johnson13}). If $Q$ is too large, the escape is limited by the surface source rate which we describe below. 

\subsubsection{\textbf{$\dot{M_S}$} Source-limited escape: thermal evaporation of an Exo-Io}\label{destruction}
Due to the extreme tidal heating and irradiation, thermal escape at an exo-Io may also contribute to the total Na I \& K I mass loss. Based on the surface T$_0$, we compute two possible surface source rates as at Io: $\dot{M}_{0,evap}$ and $\dot{M}_{0,volc}$. 
Improving upon previous calculations (Eqn. 36: \citet{cassidy09} derived from \citet{fc87} ), we use the experimentally-derived values of \citet{vanlieshout14},based on an Arrhenius-type vapor pressure relation: P$_{vap} = exp(-A/T_0 + B)$ (Eqn. 13 \citet{vanlieshout14}), where we use the average of two rocky mineral end members, enstatite MgSiO$_3$: A= 68 908 $\pm$ 8773 K;  B = 38.1 $\pm$ 5.0 and fayalite Fe$_2$SiO$_4$: A= 60 377 $\pm$ 1082; B= 37.7 $\pm$ 0.7. The source rate due to thermal evaporation can be written as:   
\begin{equation}\label{mdot0evap}
\dot{M}_{0, evap} = x_i 4 \pi R_s^2 P_{vap} (T_0) \left (\frac{m_i}{2 \pi k_b T_0} \right )^{1/2} 
\end{equation}

where m$_i$ is the mass of the volatile atom or molecule in question and $x_i$ is the mass fraction. We use a lower limit to the chondritic composition of Io constrained by \citet{fegleyzolotov2000} to be $x_{Na} = 0.05$. Based on the variety of Na/K ratios in Table \ref{alkalis}, it is conceivable that $x_{Na}$ will vary over time and further geophysical modeling assessing interior-atmosphere coupling is needed (e.g. \citet{noack17}; \citet{Bower19}). \\

The tidally-limited volcanic source rate for an exo-Io can be roughly estimated as $\sim \eta m_i \frac{\dot{E_s}}{ U_{s}}$ based on Eqn. \ref{tidalheating} with an efficiency $\eta$. In Section \ref{tidally-driven}, based on direct observations of the volcanic species NaCl and SO$_2$, we validated the tidally-driven Na I supply at Io. Directly scaling to this supply, albeit variable, we can write the tidally-driven volcanic source rate at an exo-Io as: 
\begin{equation}\label{mdot0volc}
\dot{M}_{0, volc: Exo-Io} \sim \eta_{\mathcal{T}} \dot{M}_{0, volc:Io}
\end{equation}

The total source rate, like at Io, is evaporative and volcanic : $\dot{M_0} \sim \dot{M}_{0, evap} + \dot{M}_{0, volc}$. Therefore, the source-limited escape rate due to surface heating for a given species of surface-Jeans parameter $\lambda_0 = \frac{G m_s m_i} {R_s k_b T_0}$ can be written as: 
\begin{equation}\label{dmdtS}
\dot{M}_{S} = \mathcal{R} \dot{M_0} (1+ \lambda_0) exp(-\lambda_0) 
\end{equation}

where $\mathcal{R}$ depends on $n_0$ the surface density of the species following Eqn. 2A in \citet{johnson15} \footnote{ $\mathcal{R} = \left ((n_0 \sigma R_s)^{-0.09} + (0.014n_0 \sigma R_s \lambda_0 ^{2.55} exp(-\lambda_0))\right ) ^{-1}$} where $\sigma_i$ is the collisional cross section of the escaping species in question, typically $\sim 10^{-15}$cm$^{2}$ for N$_2$ in the molecular kinetic simulations from \citet{volkovjohnson13} to which the parameter is fit to. This estimate is in principle more accurate than the standard Jeans escape used in the literature as it includes the suppressed escape due to Fourier conduction as demonstrated in recent molecular kinetic simulations of volatile escape at small bodies ( \citet{volkov11}; \citet{volkovjohnson13}; \citet{johnson15}). In Table \ref{T5exoio} we consider the evaporation timescale of the entirety of an exo-Io (composed of MgSiO$_3$ - Fe$_2$SiO$_4$) by focusing on only the \textit{total} source-limited escape described above, as a first approximation.


\begin{table*}[t]
    \centering
    \hspace*{-1.5cm}
\begin{tabular}{lrrrrrrrrrrrr}
\hline
Alkaline System  &  L$_{XUV}$ [ergs/s] & $f_{mag}$ & $f_{ram}$ &   $\beta$ &   $\eta_{\mathcal{T}} $ &   T$_{0}$ [K] &   H$_{0}$ [km] &   $\lambda_{0, Na}$ &   $\mathcal{R}_{SO_2}$ &   R$_{x}$ [R$_{Io}$] & a$_{Hill, s}$ [R$_{Io}$] \\ 
\hline
 Io  &     1.4 $\times$ $10^{28}$  &                        1.0 &               1.0 &               1.0  &  1.0 & 110  &         8.3 &                83 &          10$^{-13}$    &       1.25 & 5.8 \\
 WASP-52b  &     3 $\times$ $10^{29}$ &   2 $\times$ 10$^{4}$             &        1.8 &               27 &               9.5 $\times$ $10^{5}$  &       2211 &         160 &                4.1 &            10$^{-7}$ &       3.2 & 1.9 \\
 WASP-76b  &       8 $\times$ $10^{27}$ &   3 $\times$ 10$^{4}$             &                     0.5 &              243 &                    8.1 $\times$ 10$^{5}$ &       3049 &         220 &                3.0 &            10$^{-11}$ &       4 & 2\\
 HD189733b &    $10^{28}$ &      5 $\times$ 10$^{2}$              &                      1.9 &               6 &                  2.9 $\times$ 10$^{5}$  &       1866 &         135 &                4.9 &            10$^{-5}$ &       2.7 & 2.2 \\
 XO-2 N b  &  8 $\times$ $10^{27}$ &   6 $\times$ 10$^{2}$              &                            1.4 &               13 &                   1.3 $\times$  10$^{6}$  &       2042 &         147 &                4.4 &            10$^{-6}$ &       2.8 & 2.3\\
 WASP-49b  &  8 $\times$ $10^{27}$ &    2 $\times$ 10$^{3}$               &                         2.3 &              38 &                     9.4 $\times$ 10$^{5}$ &       1902 &         137 &                4.8 &            10$^{-5}$ &       2.6 & 2.6 \\
 HAT-P-12b &   3 $\times$ $10^{28}$  & 10$^{3}$   &                        2.2 &               9 &                      4.6 $\times$ 10$^{5}$  &       1379 &          100 &                6.6 &            1 &       2.1 & 2.9 \\
 WASP-6b   & 3 $\times$ $10^{27}$  &     10$^{3}$ &                        1.6 &               17 &                      3.6 $\times$ 10$^{5}$ &       1546 &         112 &                5.9 &            10$^{-1}$ &       2.3 & 2.6 \\
 WASP-31b  &  8 $\times$ $10^{27}$  &   3 $\times$ 10$^{3}$ &                         1 &              82 &                     3.4 $\times$ 10$^{5}$ &       1970 &         142 &                4.6 &            5 $\times$ 10$^{-6}$ &       2.6 & 3.0\\
 WASP-96b  &   5 $\times$ $10^{27}$  &     5 $\times$ 10$^{2}$  &                       1.2 &               24 &                     3.3 $\times$ 10$^{5}$ &       1672 &         121 &                5.4 &            10$^{-3}$ &       2.4 & 3.2\\
 HD209458b &   $10^{28}$  &      10$^{3}$  &                      1.1 &              43 &                    3.0 $\times$ 10$^{5}$  &       1827 &         132 &                5.0 &             5 $\times$ 10$^{-5}$ &       2.5 & 2.9 \\
 WASP-17b  &      $10^{28}$ &     10$^{4}$ &                    0.9 &              173 &                     2.1 $\times$ 10$^{5}$ &       2087 &         151 &                4.3 &            5 $\times$ 10$^{-7}$ &       2.7 & 3.1 \\
 WASP-69b  &    4 $\times$ $10^{28}$  &   9 $\times$ 10$^{2}$ &                       1.7 &               9 &                     1.8 $\times$ 10$^{5}$ &       1296 &          94 &                7.0 &            1 &       2.0 & 3.2  \\
 WASP-39b  &    8 $\times$ $10^{27}$  &   10$^{3}$ &                       1.4 &               25 &                     1.4 $\times$ 10$^{5}$&       1433 &         104 &                6.3 &            1 &       2.1 & 3.4 \\
 HAT P 1-b &   $10^{28}$ &     8 $\times$ 10$^{2}$ &                       0.9 &              30 &                      8.7 $\times$ 10$^{4}$ &       1600 &         116 &                5.7 &            10$^{-1}$ &       2.2 & 3.2 \\
\hline
\end{tabular}
    \caption{\textbf{Exoplanetary Sodium and Potassium Parameters.} Listed quantities are computed from known stellar quantities in Table \ref{T2stellar}. The XUV luminosities L$_{XUV}$ are computed from G and K stars from \citet{lammer09}. The ratio of magnetic and ram pressures at the alkaline planetary system with Io are described as the parameters $f_{mag} = \frac{P_{mag,p}}{P_{mag, Io}}$ and $f_{ram} = \frac{P_{ram,p}}{P_{ram, Io}}$  where P$_{mag, Io} = 1.5 \times 10^{-5}$ dynes cm$^{-2}$ and P$_{ram, Io} = 2.4 \times 10^{-6}$ dynes cm$^{-2}$. The canonical ratio of radiation pressure acceleration to gravity, $\beta$, is estimated using the parameters in Table \ref{T2stellar}. The satellite $\beta_s$ is additionally defined in Eqn. \ref{Eqnradpressure}. The tidal efficiency, $\eta_{\mathcal{T}}$ ,  with respect to Io is the ratio of tidal dissipation rates $\dot{E}_s/ \dot{E}_{Io}$ as in Eqn. \ref{tidalheating}. The net surface temperature including tidal heating is T$_0$, as in Eqn. \ref{tidalenhancement}. The scale height evaluated at T$_0$ for the volcanic volatile SO$_2$ is listed as H$_0$. The Jeans parameter for Na is $\lambda_0$. The conduction prefactor, $\mathcal{R}_{SO_2}$, describes suppressed surface Jeans escape of the entire atmosphere as in Eqn. \ref{dmdtS}. The exobase radius as defined in Eqn. \ref{Eqn:Rx}, R$_x$, is in units of Io radii: R$_{Io} = 1822$ km. As described in the text, the exobase for Na is derived using the observed SO$_2$ density at the surface n$_{0, SO2} = 1.8 \times 10^{11}$ cm$^{-3}$ \citep{lellouch15} and mixing ratio, X$_{NaCl} = 0.013$ \citep{lellouch03}, based on venting from the surface.  
    } 
    \label{T4sodiumpotassium}
\end{table*}

Finally, we estimate t$_{evap} \sim \frac{M_{Io}}{\dot{M}_{S, tot}}$ the critical timescale to evaporate the entirety of the exo-Io. As a rough bound to the silicate component of the exo-Io we assume the total loss eroding the surface is $\dot{M}_{S, tot} \sim \dot{M}_{S}$ where  $x_i = 1.0$.
As the evaporation timescale describes the entirety of the exo-Io describing the silicate evaporation of MgSiO$_3$ \& Fe$_2$SiO$_4$, we consider only source-limited escape (Eqn.12 improved from Jeans escape). We find that for several candidate systems less than 1\% of an exo-Io would have evaporated throughout the stellar lifetime (t$_*$: Table \ref{T2stellar}) which we have boldfaced based on our range of mass loss rates. We flag potentially disintegrated exo-Io systems from our mass loss model as \textcolor{red} {$\oslash$} as well as the $T_0 \gtrsim 2000$ K catastrophic disintegration from \citet{perez-becker13} as \textcolor{red} {$\bigcirc$}. The large uncertainties in stellar age permit the possibility that several of these flagged system still have an Io-mass satellite today. While we have modeled the principal mass loss processes, these estimates should only be used as a guide as more detailed modeling is needed to assess the fate of these systems. That is, we have assumed the rheological properties of the moons are similar to Io's yielding the tidally-heated surface temperatures (Table \ref{T4sodiumpotassium}:T$_0$) dominating the putative satellite destruction. Figure \ref{fig1} additionally marks the critical destruction limit for Sun-like systems using Eq. \ref{tidalenhancement} at $T_0 = T_{eq}+ \Delta T_0$ (yellow) and $T_0 = T_{eq}$ (red).

Satellite destruction implies a large source of circumplanetary material possibly in the form of a planetary ring (e.g. \citet{burns99}) or gas torus (\citet{johnsonhuggins06}) capable of generating alkaline signatures due to the extreme photodesorption. Saturn's stable toroidal atmosphere of O$_2$ is supplied purely by the photodesorption and sputtering of ice grains in its planetary ring \citep{johnson06a}. A lower limit to desorption from such debris is provided in Table \ref{T5exoio} as $\langle \dot{M_0}\rangle_{d}$, which also serves as a lower limit for a desorbing Trojan source with negligible tidal heating. Dust signatures from a catastrophically disintegrating body appear to be rare as is the case for the planetary systems at the anomalous stars KIC12557548, KIC 8462852, RZ Piscium and WD 1145+017 (\citet{rappaport2012}; \citet{boyajian16}; \citet{punzi18}; \citet{vanderburg15}). Nonetheless, the fact that \textit{all} strongly outgassing alkaline systems have sub-Saturnian densities (Table \ref{T3planet}:$\ominus$) may suggest a close-in torus of exogenic material \citep{exorings}.

As a final step we take the net mass loss rates $\dot{M}_{Exo-Io}$, and translate them to the Na I supply rates for the exoplanetary system, and compute the line-of-sight column densities averaged over the entire stellar disk (Eqn.\ref{eqn:Ncolumn}) shown in Table \ref{T5exoio}:

\begin{equation}\label{Eqn:exoio_column}
\langle N \rangle _{Exo-Io}= \frac{\frac{\langle \dot{M}_{Exo-Io} \rangle}{m_i} \tau_i}{\pi R_* ^2} 
\end{equation}

$\dot{M}_{Exo-Io}$ is the dominant mass loss rate based on the three mechanisms we compute: source-limited $\dot{M}_{P}$ (Eqn. \ref{dmdtP}), energy-limited $\dot{M}_{P}$ (Eqn. \ref{dmdtP}), and plasma-limited $\dot{M}_{P}$ (Eqn. \ref{dmdtP}) in Table \ref{T5exoio}. Therefore, the total number of absorbers $\mathcal{N}$ along a LOS, is limited by the alkaline lifetime so that: $\mathcal{N} \sim \frac{\dot{M}_{Exo-Io}}{m_{i}} \tau_{i}$, where $\tau_{i}$ is assumed to be limited by ionization. The line-of-sight column against the stellar disk is then simply $ N_{Io} \sim \frac{\mathcal{N}}{\pi R_*^2 } $. The average LOS column densities we estimate in Table \ref{T5exoio} can exceed both Jupiter's atmospheric column density $\sim 7 \times 10^{11}$cm $^{-2}$ as well as Jupiter's $\sim 1000 R_{J}$ exospheric column $\sim 2.5 \times 10^{10}$cm $^{-2}$ (Section \ref{exogenicsources}), by three orders of magnitude.

These range of predicted exo-Io column densities can be directly compared to the minimum required column density derived from the observed equivalent widths from high-resolution transmission spectra observations which we will now describe.

\subsection{Sodium and Potassium Gas Column Densities: Constraints from Observations}\label{columnsection}
Despite considerable advances in instrumentation and advanced techniques to probe exoplanet atmospheres with line profiles, the geometrical distribution of the Na I \& K I seen in transit cannot be inferred. Given that Na I and K I line cores are capable of probing extremely tenuous pressures, the optically-thin regime can illuminate a \textit{minimum} line-of-sight column density: 
\begin{equation}\label{Eqn:obscolumn}
\langle N \rangle _{obs}= \frac{W_{\lambda,i}} {\frac{\pi e^2}{m_e c^2} f_{ik} \lambda^2} 
\end{equation}
\citep{draine2011}, where $f_{ik}$ is the oscillator strength, for the Na D1 and D2 lines and the equivalent width in angstroms W$_{\lambda,i}$. We can now compare this observed column density to our estimates of an exo-Io at a hot Jupiter based on our understanding of alkaline mass loss. The comparison of required column densities indicates that a large majority of the systems are consistent with an exogenic supply of Na I atoms such as an exo-Io. Based on these rough extimates, the supply of exogenic Na I appears to be dominated by atmospheric sputtering ($\dot{M}_P$) for six systems and by energy-limited escape ($\dot{M}_U$) for six systems, with only two systems by escape due to surface heating ($\dot{M}_S$). The maximum supply, considering all upper limits, is underlined in Table \ref{T5exoio}. We discuss limitations and additional plasma interactions in the following section in the context of observing and characterizing these signatures with upcoming instrumentation. \\

\begin{sidewaystable*}
    \centering
    \vspace*{7.5cm}
\begin{tabular}{lrrrrrrrrrrr}
\hline
Exo-Io &   $\langle \dot{M_0} \rangle_{evap}$ &   $\langle \dot{M_0}
\rangle_{volc}$ &   $\langle \dot{M_0}\rangle_{d}$ &   $\langle \dot{M_{S}}\rangle$&   $\langle \dot{M_{U}}\rangle$ &   $\langle \dot{M_{P}}\rangle$  &  $^{\textcolor{violet}{\star}}\langle N\rangle_{K, Exo-Io}$ & $^{\textcolor{red}{\bullet}}\langle N\rangle_{Na, Exo-Io}$  &   $\langle N \rangle _{Obs}$    & $\langle t \rangle_{evap}$ & $ t_{*}$  \\
\hline
  &  [kg/s]  & [kg/s] & [kg/s] &[kg/s] &[kg/s]  &[kg/s]      &   [K cm$^{-2}$] & [Na cm$^{-2}$]   & [Na cm$^{-2}$]    &    [Gyr] & [Gyr] \\
\hline
 \textcolor{red} {$\bigcirc$} WASP-52 b I  & $10^{12.1 \pm 0.8}$ &    $10^{9.4\pm 1}$ &                                       $ 10^{3.6\pm 1.4}$ &                                       $10^{2.6\pm .1 }$ &  \underline{$10^{6.8\pm .2 }$}     &                               $10^{4.7 \pm 1.9}$ &                           $10^{11.2\pm 1}$       &                                 $10^{13\pm 1}$ &                                  $10^{11.5\pm 0.05}$ &                                                                                                    360 $\pm $ 60 & 0.4 $\pm$ 0.3 \\
 
  \textcolor{red} {$\bigcirc$} WASP-76 b I & $10^{15.5\pm 0.5}$ &    $10^{9.3\pm 1}$ &                                       $10^{12\pm 0.8}$ &                                       $ 10^{2.9 \pm .1}$ &                                    \underline{$10^{5.1\pm 2.6}$} &                                  $10^{4.7 \pm 2.2}$ &       $10^{7.8\pm .6}$ &                           $10^{9.6\pm .6}$ &                                  $10^{10.1\pm 0.05}$ &                                                                                                    180 $\pm $ 35 & $\sim$5 \\
 \textbf{HD189733 b I} & $ 10^{9.8\pm 1}$ &    $10^{8.9\pm 1}$ &                                       $10^{1.6\pm 1.5}$ &                                       $10^{2.8\pm .3}$ &                                    $10^{5.4\pm 0.2}$ &                                  \underline{$10^{4 \pm 1.2}$} &      $10^{9.8\pm .5}$ &                                $10^{11.6\pm .5}$ &                                 $10^{10.4\pm 0.05}$ &                                                                                                230  $\pm $ 110 & 4.3 $\pm $ 2.8\\
 \textcolor{red} {$\bigcirc$} XO-2 N b I  & $10^{11.1\pm 0.9}$ &    $10^{8.5\pm 1}$ &                                       $10^{6.2\pm 1.2}$ &                                       $10^{2.5 \pm .1}$ &                                    $10^{5\pm 0.2}$ &                               \underline{$10^{4 \pm 1.3} $} &   $10^{8.6\pm .5}$ &                                $10^{10.5\pm .5}$ &                                $10^{11\pm 0.05 }$ &                                                                                                    430.  $\pm $ 64 & 6.3 $\pm $ 2.4\\
 \textbf{ WASP-49 b I} & $10^{10.1\pm 1}$ &    $10^{8.4\pm 1}$ &                                       $10^{4.9 \pm 1.3}$ &                                       $10^{2.5\pm .1}$ &                                    10$^{5\pm 0.2}$ &                                  \underline{$10^{4.3 \pm 1.5}$} &            $10^{8.6\pm .4}$ &                       $10^{10.4\pm .4}$ &                                $10^{10.7\pm 0.05}$ &                                                                                                  400.  $\pm $ 30 & $\sim$5\\
 \textcolor{red} {$\oslash$} HAT-P-12 b I & $10^{4.6\pm 1.3}$ &    $10^{8\pm 1}$ &                                       $ 10^{-4.2\pm 1.9}$ &                                        \underline{$10^{5.8\pm .9}$} &                                    $10^{5.6\pm 0.2 }$ &                                 $10^{4.2 \pm 1.6 }$ & $10^{10.6\pm .7}$ &                                 $10^{13.3\pm .2}$ &                                 -- &                                                                                                   0.20  $\pm $ 0.18 & 2.5  $\pm $ 2 \\
  \textcolor{red} {$\oslash$}WASP-6 b I  & $10^{6.7\pm 1.1}$ &    $10^{8\pm 1}$ &                                       $ 10^{0.6\pm 1.6}$ &                                       $10^{4.1\pm 1.1}$ &                                    $10^{4.5\pm 0.2}$ &                                  \underline{$10^{4.2 \pm 1.7}$} &     $10^{9\pm .3}$ &                               $10^{11.1\pm .1}$ &                                 -- &                                                                                                 12  $\pm $ 10 & 11  $\pm $ 7 \\
 \textbf{WASP-31 b I} & $10^{10.6\pm 0.9}$ &    $10^{7.9\pm 1}$ &                                       $10^{7.1\pm 1.2}$ &                                       $10^{2.5\pm 0.1 }$ &                                    $10^{4.8\pm 0.2}$ &                                  \underline{$10^{4.3 \pm 1.9}$} &    $10^{8.1\pm .7}$ &                               $10^{9.9\pm .7}$ &                                 -- &                                                                                                    470  $\pm $ 70 &  $\sim$5\\
WASP-96 b I & $10^{8.1\pm 1.1}$ &    $10^{7.9\pm 1}$ &                                       $10^{3.1 \pm 1.4}$ &                                       $10^{3.1 \pm .6}$ &                                    $10^{4.6\pm 0.2}$ &                                  \underline{$10^{4 \pm 1.6}$} &    $10^{8.3\pm .5}$ &                              $10^{10.3\pm .5}$ &                                 \textcolor{red}{$<$} $10^{11.7\pm 0.05}$ &                                                                                                  110  $\pm $ 84 & 8  $\pm $ 8 \\
 \textbf{ HD209458 b I} & $10^{9.5\pm 1.0}$ &    $10^{7.9\pm 1}$ &                                       $10^{3.4 \pm 2.4}$ &                                       $10^{2.5\pm 0.1 }$ &                                     $10^{5\pm 0.2 }$ &                                 \underline{$10^{3.8 \pm 2.7}$} &  $10^{8.6\pm 1}$ &                                 $10^{10.4\pm 1}$ &                                 $10^{9.7\pm 0.05}$ &                                                                                                  410  $\pm $ 56 & 3.5  $\pm $ 1.4\\
 \textcolor{red} {$\bigcirc$} WASP-17 b I  & $10^{11.4\pm 0.9}$ &    $10^{7.7\pm 1}$ &                                       $ 10^{8.7\pm 1.1}$ &                                       $10^{2.5\pm .1}$ &                                     $ 10^{5\pm 0.2}$ &                                 \underline{$10^{4.6 \pm 2.2}$} &  $10^{8.3\pm .8}$ &                                $10^{10.1\pm .8}$ &                                $10^{10.8 \pm 0.05}$ &                                                                                                      422.  $\pm $ 80 & 3  $\pm $ 2.6\\
  \textcolor{red} {$\oslash$}WASP-69 b I & $10^{3.3\pm 1.4}$ &    $10^{7.6\pm 1}$ &                                       $ 10^{-4.1\pm 1.9 }$ &                                       $10^{5.9 \pm .3}$ &                                    \underline{$10^{5.6 \pm 0.2}$} &                                  $10^{4 \pm 1.7}$ &     $10^{10.3\pm 1}$ &                              $10^{13\pm .1}$ &                                 $10^{10.7 \pm 0.05}$ &                                                                                                    0.18 $\pm $ 0.08 & 2  $\pm $ --\\
 \textcolor{red} {$\oslash$} WASP-39 b I  & $10^{5.3\pm 1.3}$ &    $10^{7.5\pm 1}$ &                                       $ 10^{-0.1\pm 1.7}$ &                                       $10^{4.9\pm 1.1}$ &                                     $10^{4.8\pm 0.2}$ &                                 \underline{$10^{4.2 \pm 1.8}$} &     $10^{9.3 \pm .4}$ &                            $10^{11.7 \pm .2}$ &                               $10^{9.7 \pm 0.05}$ &                                                                                                      1.8  $\pm $ 1.7  &   $\sim 5 $ --\\
 \textbf{ HAT-P-1 b I} & $10^{7.3\pm 1.1}$ &    $10^{7.3\pm 1}$ &                                       $10^{3.7\pm 1.4}$ &                                       $10^{3.2 \pm .7}$ &                                     $10^{4.8\pm 0.2}$ &                                 \underline{$10^{4.1 \pm 1.6}$} &     $10^{8.6\pm .8}$ &                              $10^{10.6\pm .8}$ &                                $10^{11.2 \pm 0.05}$ &                                                                                                    91  $\pm $ 74  & 3.6  $\pm $ --\\
\hline
\end{tabular}
    \caption{\small \textbf{Candidate Exo-Io Sodium \& Potassium Mass Loss Calculations.} Order of magnitude estimates to assess the plausibility of an exo-Io source at each alkaline exoplanetary system. $\langle \dot{M_0} \rangle_{evap}$ (Eqn. \ref{mdot0evap} ) and  $\langle \dot{M_0} \rangle_{volc}$ (Eqn. \ref{mdot0volc}) are the Na I source rates at T$_0$ (Eqn. \ref{tidalenhancement}) due to thermal evaporation and volcanism respectively. For thermal evaporation we consider a range of mineral vapor pressures bounded by MgSiO$_3$ and Fe$_2$SiO$_4$. For tidally-driven volcanism our results are scaled to Io for a range of tidal efficiencies $\eta_{\mathcal{T}}$ corresponding to $\approx 2$ dex in rheology parameters. $\langle \dot{M_0} \rangle_{d}$ is a lower bound to thermal evaporation (Eqn. \ref{dmdtS}) evaluated at the equilibrium temperature T$_{eq}$ with a nul value of tidal heating $\dot{E}_s = 0$ (Eqn. \ref{tidalheating}). The subsequent mass loss ranges  $\langle \dot{M_{S}}\rangle$ (Eqn. \ref{dmdtS}),  $\langle \dot{M_{U}}\rangle$ (Eqn. \ref{dmdtU}),  $\langle \dot{M_{P}}\rangle$ (Eqn. \ref{dmdtP}) are provided for completeness considering 3 principal mechanisms: surface heating, upper-atmospheric heating, and plasma heating respectively. $\langle \dot{M_{S}}\rangle$: Source-limited escape is directly limited by $\langle \dot{M_0} \rangle $ and is strongly dependent on conduction to the surface as described by $\mathcal{R}$, as provided in Table \ref{T4sodiumpotassium}. $\langle \dot{M_{U}}\rangle$: Upper-atmosphere/Energy-limited escape depends on the incident L$_{xuv}$ radiation at the star derived from Lammer et al. as in \citet{jinmordasini18}, where we consider heating efficiencies between 0.1-0.4. Due to the extreme irradiation we note that it is likely that this rate is an upper limit (\citet{johnson13}). $\langle \dot{M_{P}}\rangle$: Plasma-limited escape depends strongly on the exobase radius, and the incident plasma pressure $P_{s}$ where our lower bound is stellar-wind limited and an upper bound including the magnetic pressure of the gas-giant (see Table {\ref{T4sodiumpotassium}} and c.f. Table 4.1 \citet{bob90book}). The maximum mass loss rates are \underline{underlined} based on the upper limits. The predicted average exo-Io column density based on the principal mass loss source for Na I (\textcolor{red}{$\bullet$}) and K I (\textcolor{violet}{$\star$}) is provided as: $\langle N\rangle _{Exo-Io}$ (Eqn. \ref{Eqn:exoio_column}) and directly compared to the derived column from transmission spectroscopy observations  $\langle N \rangle _{Obs}$ (Eqn. $\ref{Eqn:obscolumn}$). The range of values from transmission spectroscopy is tightly constrained as the equilibrium optical depth is fundamentally $\tau \sim 1$, although the equivalent width can indeed vary. Systems where $\langle N \rangle _{Exo-Io}$ \textcolor{red}{$<$} $\langle N \rangle _{Obs}$ are flagged as problematic for an exogenic lunar source. The evaporation timescale for the Exo-Io $t_{evap}$ is estimated based on our knowledge of the surface loss rate $\langle \dot{M_{S}}\rangle$ and compared to the system age $t_{*}$. Potentially disintegrated debris systems are flagged as \textcolor{red}{$\oslash$} from our mass loss model, whereas \textcolor{red}{$\bigcirc$} indicates that the tidally-heated T$_0 > $2000K may imply catastrophic disintegration (\citet{perez-becker13}). Following analysis, the top exomoon systems are \textbf{boldfaced}. }     

    \label{T5exoio}
\end{sidewaystable*}

\section{Discussion}\label{discussion}
\begin{figure*}\label{exoarch}
  \includegraphics[width=\textwidth]{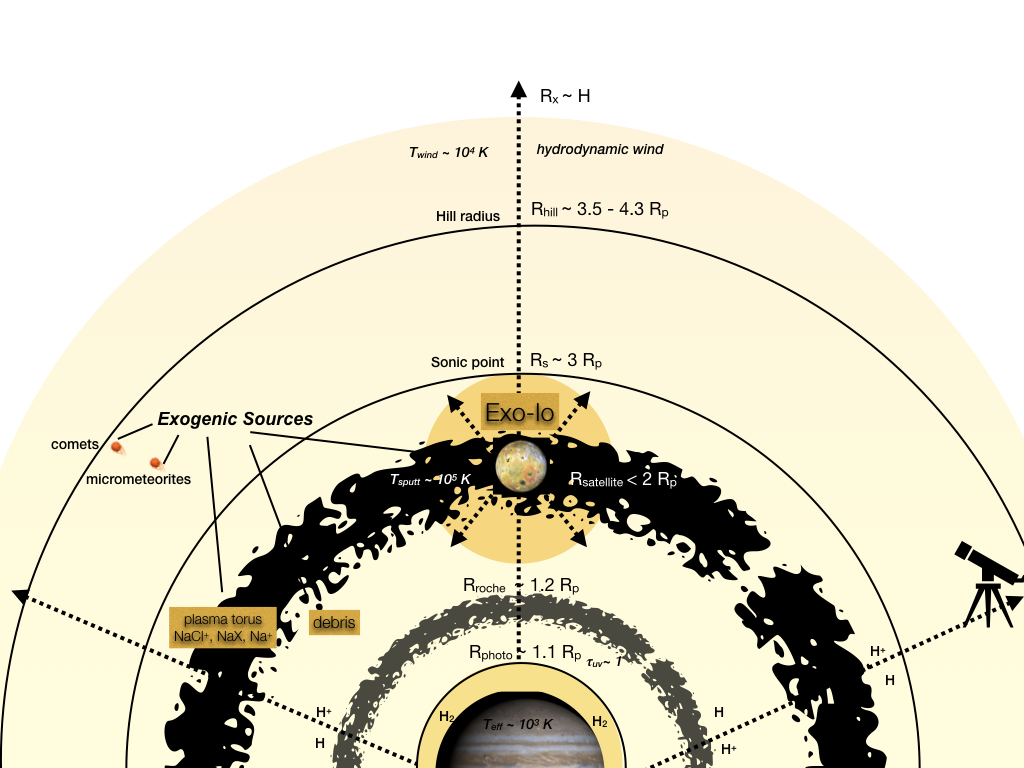}
  \caption{\textbf{2-D ``Birds-Eye'' Architecture of a Sodium Exosphere at a Close-in Gas Giant Exoplanet System.} \textit{Endogenic:} An extended sodium layer above the hot Jupiter is shown to illustrate a well-mixed Na I component (yellow). \textit{Exogenic:} An exo-Io sodium cloud (yellow) is shown at $\sim 2 R_p$. Parameters for an escaping atmosphere are overlayed, inspired by the 1-D hot Jupiter atmosphere by \citet{mc09}. Several other desorbing-exogenic sources (comets, micrometeorites, debris) illustrate the possible endogenic-exogenic interaction contributing to the light yellow exosphere. A full Monte Carlo simulation is likely necessary to describe the precise ion-neutral and electron-neutral interactions in detail, yet a upper atmosphere model by \citet{huang17} along with high-resolution observations of HD189733b (\citet{wyttenbach15}) and WASP 49-b (\citet{wyttenbach17}) are suggestive of the parameter space. If the hot Jupiter is magnetic, a slowly rotating plasma torus carrying ejected material (in black) should be present. Given that an active satellite would be orbiting close-in, orbiting at $\sim 20 km/s$ with ejected sodium speeds $\sim 10$ km/s due to sputtering (i.e. Io), the corresponding kinetic temperature will be on the order of 10$^5$ K, far hotter than the escaping hydrodynamic wind of the gas giant providing a source for line broadening. The ionized gas H, H+ above the photoionization base should also contribute strongly to the plasma density (n$_e \sim 10^9$) possibly resulting in stray sodium spraying throughout the system and into the exosphere.    }
\end{figure*}

The activity occurring on close-in, irradiated exomoons, as described here, appears to be capable of delivering a significant volatile mass to its host exoplanetary system.  The principal question which remains is how significant is the exogenic mass when compared to the expected endogenic mass, and is this additional mass already detected in the robust transmission spectra observations of Na I and K I? Nominal mass loss models from hot Jupiters estimate $\sim 10^{7}$ kg/s of total mass corresponding to roughly $\sim 10$ kg/s of Na for the nominal solar abundance scenarios whereas our exo-Io model can supply on the order of $4 \times 10^{6}$ kg/s of pure Na I, approaching the maximum volcanic output of Io and in effect, the possible destruction limit of an exo-Io. Section \ref{jupitersNa} helped qualify the mass of exogenic Na I probed in transmission at a cold Jupiter. For hot Jupiters more comprehensive searches of volatiles and their spectral imprint with high-resolution \'{e}chelle spectrographs will certainly improve our understanding of endogenic-exogenic interactions. In the last decade our early understanding of a hot Jupiter's environment was limited to H I and Na I detections at HD209458b \citet{charbonneau02}, and was largely described in the seminal paper on atmospheric escape from hot Jupiters (c.f. \citet{mc09} Figure 8 summarizing the environment in 1-D). Since then a series of observations may have enhanced our picture.

\subsection{A 2-D View of Atmospheric Escape: Plasma Tori}
In addition to the dozens of alkaline detections, several other extrasolar volatile species, He I (\citet{he2018}; \citet{spake2018}), Mg I \citet{mgvidal13} and also the ions Ca II (\citet{ca}; \citet{Ridden-Harper2016}), Fe II, Ti II (\citet{jens2018}), along with tentative detections of doubly ionized species Mg III \citep{mg2} have been observed ; reminiscent of not only Io's volatiles but also Mercury's where the stellar wind interaction is well-observed and well-simulated (e.g. \citet{killen2001}; \citet{leblancjohnson10}; \citet{schmidtPHD}). The magnesium ions were observed in the ultraviolet, where anomalies in the ingress and egress transit spectra at HD189733b and WASP 12-b have already tentatively suggested a plasma torus  (\citet{mg2}; \citet{kislyakova16}; \citet{bj2014}) as predicted in general by \citet{johnsonhuggins06}. A plasma torus is a natural consequence when volatiles emanate from an orbiting exogenic source within the magnetosphere of a planet, as the volatiles will eventually be ionized and form a torus of its ejected material. The possibility of a plasma torus adds a 2nd dimension to the hot Jupiters 1-D environment described previously, which we illustrate in black in Figure \ref{exoarch}. This 2-D illustration summarizes the Na I exosphere we describe in this work by our orbital stability constraints and simple modeling of an exo-Io's Na I escape. We indicate the various scale lengths following the same convention as \citet{mc09} inspired by parameters from HD209458b, whose escaping atmosphere has been shown to be consistent with the energetic neutral atoms (ENA) H$^{*}$ , O$^{*}$, and C$^{+}$* \citep{bj2010}. The dissociative recombination of NaCl$^{+}$ as mentioned in Section \ref{magnetotidal} yields the ENA Na$^{*}$ in the form of streams as shown in Fig. \ref{Nanebula}.  A unique consequence of exomoons orbiting hot Jupiters is that they would be directly embedded in the hydrodynamically \textit{escaping} endogenic medium rendering the distinction between the two difficult at present. However, the recent the discovery of Na I at a remarkably high altitude $R_{Na} \sim 1.5 R_p$ for WASP 49-b \citet{wyttenbach17} we discuss may be a strong indication of exogenic sodium, although not unambiguous at present. Further testable predictions as we have described include non-solar Na/K ratios, and spectral shifts due to radiation pressure.

\subsection{Evidence of Geologically-Active Satellites?}\label{endoexo}
\subsubsection{Radiation Pressure and Variability Signatures}
The radiation pressure on any emitted alkaline atom at hot Jupiters is significant, roughly $\sim 10 \times$ that at Mercury and may lead to an observable spectral shift $\Delta v_{rad}$ analogous to Mercury's comet-like Na I and K I tails (\citet{schmidt13} ; \citet{schmidt18lpi}). We compute the acceleration due to radiation pressure on a Na I atom following \citet{chamberlain61}, which can be understood by the commonly used parameter $\beta$ as the ratio of accelerations due to radiation pressure and gravity. For an orbiting atom the latter is $\frac{v_{orb}^2}{a_s}$ so that the parameter can be written as $\beta_s$:

\begin{equation}\label{Eqnradpressure}
\beta_s =  \frac{ \mathcal{F}_{\lambda} \frac{\pi e^2 \lambda^2}{m_{Na} m_e c^3} \left ( f_{D1} + f_{D2} \right )  }{\frac{v_{orb}^2}{a_s}}  
\end{equation}

where $\mathcal{F}_{\lambda}$ is the incident stellar radiation flux \citep{rybickiandlightman} at the doublet wavelength $\lambda_{Na D1,D2}$ and f$_{NaD2} = 0.641$, f$_{NaD1} = 0.320$ the oscillator strength of the Na D2 and D1 resonance lines respectively \citep{draine2011}. In Table \ref{T4sodiumpotassium} we give the value of $\beta$ for each exoplanetary system, where the stellar spectrum for an F, G, and K is used. To appreciate this effect, given that $\sim 50.7 km/s$ corresponds to a shift of 1 $\AA$, the linear velocity shift on a stagnant Na I cloud due to radiation pressure alone is then $\Delta v_{rad} \sim a_{rad} \tau_{Na}$, resulting in shifts between 10-70 km/s ($\sim 0.2 - 1.4 \AA$). Of course if the cloud is vented from an orbiting satellite, the observed shift would be $\pm \gtrsim 20$ km/s depending on both the movement of the ejected gas and the satellite's orbital motion during the observation. Accounting for these additional motions, one may expect Na I and K I clouds to move with respect to the planet's rest frame velocity on the $\sim$ km/s level. Since a molten exo-Io will additionally thermally evaporate material asymmetrically at the subsolar hemisphere, it is conceivable that spectral shifts due to radiation pressure could be strongly mitigated. Furthermore, at HD189733b the leading and trailing limbs ($-5.3\substack{+1.0 \\ -1.4}$\,km/s (blueshift) $+2.23\substack{+1.3 \\ -1.5}$\,km/s (redshift) respectively) \citet{loudenwheatley15} have not been attributed to radiation pressure but rather eastard equatorial jets, motivated by atmospheric circulation models. On the other hand,  HD209458b (-14.7 km/s), HAT-P-1b (-3 km/s), Kelt 20-b (26 km/s), HD 80606b (-4 km/s) and possibly the super earth 55 Cancri-e (-27 km/s) also exhibit spectral shifts of Na I and K I (\citet{Snellen2008}; \citet{Wilson2015}; \citet{Colon2012} ; \citet{Ridden-Harper2016}) suggesting time-sensitive monitoring of these systems is required. This interaction implies there would be considerable variability in the Na I \& K I lines, an effect observed for WASP 31-b, attributed so far to instrumental effects \citep{gibson19}. Ingress and egress observations probing the entirety of the planetary Hill sphere would also help constrain possible variability.

\subsubsection{Sodium/Potassium Ratios: Lessons from the Solar System}
As Fig.\ref{Nanebula} depicts, Io's Na I jets can travel the entirety of Jupiter's magnetosphere. Therefore whether Io was the exogenic source of the Na I discovered at Europa \citet{brown96} was uncertain until the discovery of K I \citet{brown01}. As Na/K ratios are suggestive of physical processes, the theoretical Na/K ratio at Europa by \citet{leblanc02} and (\citet{johnson02} Table I) when compared to the observational Na/K ratios at Io and Europa, demonstrated that Europa's Na I was primarily an endogenic ocean source. As we have no knowledge of the geological composition of the exo-Ios we describe in this work, we can only say with confidence that the ratios should be non-solar as tabulated in Table \ref{alkalis}. Given that current working models of Na I at hot Jupiters use solar abundance based on the early atomic gas phase predictions for brown dwarfs (\citet{seagersasselov2000}; \citet{sudarsky2000}), the observations of Na/K should roughly follow suit allowing for difference in the volatility and the mass of Na \& K. The fact that certain alkaline planets have observed K I without a corresponding Na I signature and vice-versa at certain alkalinene planets is likely problematic for a purely endogenic source. Assuming core accretion or disk instability planet formation mechanisms for the gas giants, the metals should roughly follow solar abundance along the mass-metallicity trend (\citet{thorngren2016}, \citet{mordasini2012metals}). Despite the differences in mass and in alkaline ionization potentials, the Na/K ratios at the three exoplanets with both Na \& K are still quite unusual (e.g.,  (Na/K)$_{H1b} \sim 0.5$ along with lunar-like ratios at (Na/K)$_{XO2b} \sim 6$ \&  (Na/K)$_{W39} \sim 7$). We encourage follow-up observations of these bodies especially in K I, which is more difficult to probe than Na I, possibly explaining the non-detections. While no body in the Solar System can explain a K I enhancement relative to Na I, or the non-detections (\citet{nikolov18} retrieval results in (Na/K)$_{W96} \gtrsim 10000$) such a stark contrast could be conceivable geophysically or be indicative of an extreme case of mass loss.
\begin{figure*}\label{WASP49}
  \includegraphics[width=\textwidth]{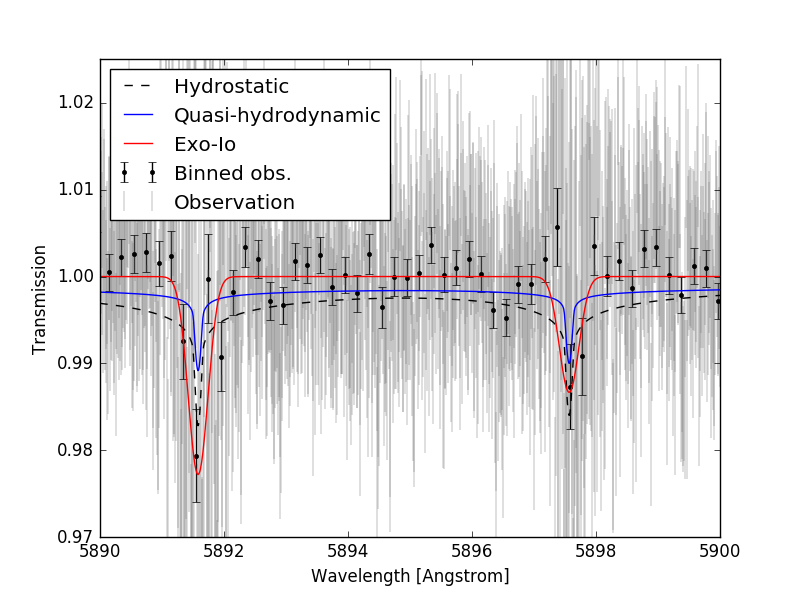}
  \caption{ Transmission spectrum of WASP 49-b at the Na doublet. We model the endogenic atmosphere with both hydrostatic and quasi-hydrodynamic assumptions. The hydrostatic model (dotted, black) is isothermal at T$\sim 2950$K equivalent to the temperature probed by \citet{wyttenbach17}. The quasi-hydrodynamic atmosphere in blue as modeled by \citet{cubillos17}, corresponds to an isothermal planetary wind of T$\sim 3000$K above a supposed isothermal T$\sim 1000$K Na$_2$S haze layer fixed at $\sim$ 100x solar abundance. We note that like HD189733b \citet{huang17} an isothermal atmosphere is unlikely. The true non-isothermal hydrodynamic solution would produce even smaller transit depth due to cooling (Fig.2 \citet{cubillos17}).  The exogenic source (red) is identical to the exo-Io of $\eta_{\mathcal{T}} \sim 10^{5 \pm 0.5}$ we describe in the text, driving a plasma-driven sodium enhancement of $\sim 10^{3.7 \pm 0.5} \times$ that at Io due to the expected stellar forcing. We employ an escaping sodium density profile (Eqn. \ref{eqnexoio}) equivalent to the 1985-1997 observations of Io's sodium corona \citep{burger01} and $\sim 10 km/s$ velocity characteristic of atmospheric sputtering. While our energy-limited escape rates also confirm a similar $> 100$X solar abundance, the planetary wind model cannot reproduce the transit depth \citet{cubillos17}. Based on the state-of-the-art 1-D hydrodynamic modeling including XUV heating, Ly-$\alpha$ cooling, dissociation, ionization and recombination by \citet{cubillos17} an endogenic atmosphere appears to be unlikely for WASP 49-b at present. We therefore find the exo-Io source to be a promising candidate for the sodium line. }
\end{figure*}

\subsubsection{An Exo-Io at WASP 49-b ?}\label{exoiow49}
The Na I transit depth was observed at the hot-Saturn WASP 49-b to extend to R$_{Na} \sim 1.5 R_p$  \citep{wyttenbach17}. This anomalously high altitude is roughly 3 $\times$ higher than HD189733b possibly warranting an alternative explanation on the origin of Na I. Furthermore, the line is significantly broader than HD189733b, suggesting a far more energetic sodium component. \citet{wyttenbach17} showed with isothermal transmission spectra models that the Na I line wings and line core can be fit \textit{individually} with two distinct temperatures (1400 K and 2950K respectively). We confirm that the transit depth can be roughly reproduced with an essentially isothermal hydrostatic atmosphere at 2950K (Fig. \ref{WASP49} hydrostatic: dotted line). The state-of-the-art 1-D hydrodynamic model developed for WASP 49-b by \citet{cubillos17} suggests however that the temperature profile derived based on the XUV heating, Ly-$\alpha$ cooling, dissociation, ionization, and recombination is not isothermal. Figure 2 of \citet{cubillos17} shows that temperatures near 3000K only persist between 0.01 - 10 nbars, after which the atmosphere significantly cools. \citet{cubillos17} Figure 8 attempted to fit the full line profile with a quasi-hydrodynamic model corresponding to two hydrostatic layers at T = 1000K and an arbitrary isothermal layer at T = 2950K escaping above an assumed Na$_2$S haze layer at $\sim$ 1 nbar at 100 $\times$ solar abundance, $X_{\odot}$. Reproducing this model (Fig. \ref{WASP49}: blue line), we confirm that it fails to reproduce the transit depth and is far too narrow as stated by the authors. The endogenic models are problematic due to the large mass of Na I required by the large transit depths. The minimum LOS columns given in Table \ref{T5exoio} can easily be converted to total mass integrated over the stellar disk, for which we find a minimum Na I mass of $M_{Na, min} > 1.3 \times 10^{6} kg$ or 10 $\times$ Jupiter's atmospheric Na I (Section \ref{exogenicsources}). As hot Jupiters' mass loss are dominated by energy-limited escape (Eqn. \ref{dmdtU}) for an incoming EUV flux of F$_{EUV} = 2500$ ergs/s/cm$^2$, and a range of $\eta_{XUV}$ described above, leads to a \textit{total} mass loss rate of $\dot{M}_{tot, W49b} \sim  10^{7.1 \pm 0.3} kg/s$. Assuming solar abundance, the Na I mass loss rate of $\dot{M}_{Na, W49b} \sim 10^{1.4 \pm 0.3} kg/s$ corresponds to a \textit{total} mass of only $M_{Na, escape} \sim 10^{3.8 \pm 0.3} kg$. This implies that the Na I abundance, if endogenically escaping, must be $X_{Na} \sim 5000 \times X_{\odot}$. A strongly super-solar metallicity is consistent with the calculations of \citet{cubillos17} yet such a haze model has difficulty fitting the transit depth. It is possible that at such high altitudes local thermodynamic equilibrium (LTE) breaks down, and a non-LTE distribution could better fit the transit depth as suspected by recent atmospheric retrieval modeling by \citet{fisher_heng2019}; however the precise broadening mechanism is as of yet unidentified. \\

Using the highest quality observations of the Na I column abundance in Io's corona from 1985-1997 by \citet{burger01}, we test the presence of an orbiting exo-Io (Fig. \ref{WASP49}: red line) based on the parameters in Table \ref{T4sodiumpotassium} and \ref{T5exoio}. Assuming the observations can be scaled and then extrapolated inwards to the surface, a tidally-heated, ionization-limited radial profile of an exo-Io consistent with the power law for Io's Na I corona would be:
\begin{equation}\label{eqnexoio}
n_{Exo-Io} (b) = \eta_{\mathcal{T}} \left ( \frac{\tau_{Na, exo-Io}}{\tau_{Na, Io}}\right ) n_{Io, 0} \, b^{-3.34} \, cm^{-3}
\end{equation}

Here b is the impact parameter in units of R$_s$, $\tau$ the Na I lifetime ($\tau_{Na, Io} \sim 4 hours$ \citet{wilson02}), and a tidal efficiency of $\eta_{\mathcal{T}} \sim 10^{5 \pm 0.5}$ in good agreement with the plasma-driven loss described in Section \ref{magnetotidal}. The column density we estimate in Table \ref{T5exoio} $\sim 9 \times 10^{10 \pm 0.4}$ is well within error of the required column density from observations.  The effective kinetic temperature of the gas we set to $\sim 1.4 \times 10^{5}$K equivalent to the canonical atmospheric sputtering distribution velocity of $\sim 10$ km/s \citet{wilson02}. We note that the line profile becomes too narrow for a thermal velocity distribution, suggesting a non-thermal velocity distribution venting from an exomoon. If our orbital stability calculation of a critical tidal $\mathcal{Q}_{p, W49b} \sim 4 \times 10^9$ for an Io-sized satellite holds against the equilibrium tide limit for hot Jupiters, one might conclude that a geologically-active satellite is a natural source for the planetary system's Na I as at our Jupiter.  \\

\section{Conclusions}
Given that the prospect of discovering extrasolar satellites with masses $M \lesssim $10 $^{-2} M_{\oplus}$ by radial velocity, or radii $R \lesssim 3 \times 10 ^{-1} R_{\oplus}$ by transit techniques are bleak at present (\citet{teacheykipping18}; \citet{kreidberg2019}), we described in this work how \textit{gas} signatures of Na I \& K I at a hot Jupiter could be indicating the presence of a geologically-active satellite subject to the ambient plasma. This appears to be the case for at least one such system,  WASP 49-b. We use the Jupiter-Io Na I system extending to $\sim 1000 R_J$ as a benchmark for our case study of an exo-Io, remarking on the fact that $\sim $ dozen exoplanetary systems host robust alkaline detections since the initial suggestion of a toroidal exoplanet atmosphere by \citet{johnsonhuggins06}. We therefore use the well-known orbital parameters of this sample to assess both the survival and possible gas contribution of an exo-Io by modeling the Na I mass loss applicable to the order of magnitude level. Working towards a general description of a rocky exomoon orbiting a hot Jupiter, we find several natural consequences.

Foremost, in Section \ref{tidalstability} we built on the calculations of massive satellites around tidally-locked close-in exoplanets in \citet{cassidy09}, showing that smaller exo-Ios can in principle survive around \textit{all} observed alkaline systems to date (Fig. \ref{fig1}) due to the gravitational forcing of the host star. If an exo-Io were to exist the canonical condition requires that the critical tidal $\mathcal{Q}$ of the planet must suffice $\mathcal{Q}_{p, crit} < Q_p$ where the upper limit of Q$_p$ is $\sim 10^{12}$ as found by \citet{goldreich77} and improved by \citet{wu05-1}. Based on the non-detection of alkalis around WASP 19-b along with the marginal detections around the ultra-short period Jupiters WASP 43-b, WASP 12-b, and WASP 103-b it is conceivable that this upper limit for hot Jupiters is closer to $\sim 10^{11}$. Although the critical orbital period we derive, $\tau_{crit} \gtrsim 0.5$ days, holds true \textit{gravitationally} over the lifetime of the planetary system, the \textit{thermal} destruction of an exo-Io is far more jeopardous based on our mass loss model of molten irradiated bodies (Section \ref{destruction}). Catastrophic disintegration due to tidally-heated surface temperatures exceeding the $\sim 2000$K threshold simulated by \citet{perez-becker13} may also suggest a critical exo-Io period beyond $\sim 2$ days (Figure \ref{fig1}).

Consequently, the Na I at the destroyed systems (Table \ref{T5exoio} ) if confirmed, would then be the volcanic remnants of a disintegrated exomoon, possibly still desorbing in the form of debris depending on the timescales. Survival of an exo-Io should therefore be considered in tandem with its mass loss, as orbital stability dependent on semimajor axis is a necessary yet insufficient condition. 
Although the stellar proximity and insolation is threatening for the above systems, we find that, for the majority of systems, the stellar tides not only force the exo-Io to remain in orbit, but also drive significant tidal heating within the satellite, roughly 5 orders of magnitude higher than Io. This significantly increases the mass flux to the planetary system, possibly contributing to the observed Na I spectra in transmission at hot Jupiters which could be of endogenic or exogenic origin. While we do not conclude that the rest of the Na I signals are solely exogenic, we cannot rule this possibility out at present. Our case for the exogenic origin of alkalis is made by our Solar System's benchmark case of Jupiter's atmospheric transmission spectrum (section \ref{jupitersNa}) for the first time, and the ability of Io's \textit{magnetospherically-driven} activity to dominate Jupiter's exospheric Na I emission (section \ref{magnetotidal}). 

Our simple estimates result in a line-of-sight column density of $\sim 7 \times 10^{11}$ Na cm$^{-2}$ for Jupiter's atmosphere. The corresponding mass arriving at Jupiter's upper atmosphere $\dot{M}_{exo} \sim$ 0.05 g/s (or $M_{Na} \sim 1.2 \times 10^{8}$ g integrated over Jupiter) we conclude can be sourced exogenically, even in excess of the required amount. In addition to Io's powerful Na I supply of $\sim$ 110 g/s, the corresponding mass from cosmic dust $\sim$ 3.7 g/s and cometary impacts  $ \sim  $ 1.3 - 9 $\times$ 10$^{2}$ g/s could also supply the Na I based on the decades of monitoring following the 1994 Shoemaker-Levy 9 impact which we walk through in detail starting in section \ref{exogenicsources}. To obtain a first estimate on how an exo-Io's Na I supply alone can influence the transmission spectrum of a hot Jupiter, we focus on the range of line-of-sight column densities which easily translate to equivalent widths in Table \ref{T3planet}, and described in Section \ref{magnetotidal}. Scaling to Io's measured atmospheric sputtering rate of $\sim 10 - 100$ kg/s of Na in Equation \ref{dmdtP}, and acknowledging the fact that an exo-Io orbiting a hot Jupiter would reduce its average Na I lifetime to $\sim 10$ minutes resulting in an occulting Na I cloud of order $\sim 1 R_J$, nevertheless yields Na I clouds 3 - 6 orders of magnitude more dense on average than at Io due to the expected close-in irradiation acting on a tidally-active body we described. The increase in density, when compared to the minimum observed column densities are well in agreement and can even be in excess of the required amount as dictated by the equivalent widths in Table \ref{T3planet}. We conclude several hot Jupiters can be sourced in principle by thermally-driven or plasma-driven Na \& K loss from an active exo-Io, boldfaced in Table \ref{T5exoio}.

Several Na I spectra appear to suggest that the Na I is dynamic either redshifted or blueshifted possibly due to radiation pressure, or broadened suggesting collisions or a non-thermal Na I distribution. Finally when compared to K I, the few planets have ratios which are largely non-solar, contrary to endogenic formation theory \citet{kreidberg2014b}.

This first work on exo-Ios \textit{hints} at the presence of hidden geologically-active satellites in several candidate systems at present, where we suggest for one such system WASP 49-b, an exo-Io may be the leading explanation. That is to date, endogenic models cannot reproduce the extraordinarily high altitude of the observed Na even with parameters from a state-of-the-art hydrodynamically escaping atmosphere. The parameters we derive for our candidate exo-Io orbiting WASP 49-b are consistent with our predictions in Tables \ref{T4sodiumpotassium} and \ref{T5exoio}, and the number density profile and corresponding velocity of Na I we fit is identical to the precise 1985-1997 observations of Io's Na I corona.

As the physical processes we describe in this work are capable of expanding material to the edge of the magnetosphere and/or Hill sphere of a Gas Giant system, it is important in this coming decade to start considering exoplanets as exoplanetary \textit{systems}. The boon of the bright sodium and K I lines may be providing astronomers the first inferences of activity from the remnants of small bodies.  

\section*{Acknowledgements}
We thank the anonymous referee for helpful comments and suggestions, in particular, in motivating an extensive alkaline metal mass loss model from an irradiated rocky exomoon.
The authors thank Phil Arras for guidance and discussion on eddy diffusion and tidal dissipation theory at hot Jupiters. A.V.O and E.L thank Kevin Heng for insightful discussions on the normalization degeneracy, Na I resonance line theory and our ability to constrain alkaline abundances today. A.V.O expresses gratitude to Pierre Auclair-Desrotour, Wade Henning, Adrien Leleu, and Sebastien Charnoz for insight on dynamic stability and discussions on tidal $\mathcal{Q}$. A.V.O also thanks Mike Skrutskie for instrumental insight on Na and K variability. REJ acknowledges grant support from the NASA PDS program. C.H is supported in part by NASA under grant program NNX16AK08G and thanks J.Steffen. A.V.O. \& C.M. acknowledge the support from the Swiss National Science Foundation under grant BSSGI0$\_$155816 ``PlanetsInTime''. Parts of this work have been carried out within the frame of the National Center for Competence in Research PlanetS supported by the SNSF.

\clearpage

\bibliographystyle{yahapj}
\bibliography{biblio_exomoons.bib}

\begin{thebibliography}{}
\providecommand\natexlab[1]{#1}
\providecommand\JournalTitle[1]{#1}

\bibitem[{{Agol} {et~al.}(2005){Agol}, {Steffen}, {Sari}, \&
  {Clarkson}}]{agol05}
{Agol}, E., {Steffen}, J., {Sari}, R., \& {Clarkson}, W. 2005,
  \href{http://dx.doi.org/10.1111/j.1365-2966.2005.08922.x}{\JournalTitle{\mnras},
  359, 567}

\bibitem[{{Alexoudi} {et~al.}(2018){Alexoudi}, {Mallonn}, {von Essen},
  {Turner}, {Keles}, {Southworth}, {Mancini}, {Ciceri}, {Granzer}, {Denker},
  {Dineva}, \& {Strassmeier}}]{alexoudi18}
{Alexoudi}, X., {Mallonn}, M., {von Essen}, C., {et~al.} 2018,
  \href{http://dx.doi.org/10.1051/0004-6361/201833691}{\JournalTitle{\aap},
  620, A142}

\bibitem[{Asplund {et~al.}(2009)}]{asplund2009}
Asplund, M., {et~al.} 2009, \JournalTitle{Annu. Rev. Astron. Astrophys.}, 47

\bibitem[{{Astudillo-Defru} \& {Rojo}(2013)}]{ca}
{Astudillo-Defru}, N., \& {Rojo}, P. 2013,
  \href{http://dx.doi.org/10.1051/0004-6361/201219018}{\JournalTitle{\aap},
  557, A56}

\bibitem[{{Barnes} \& {O'Brien}(2002)}]{barnesobrien02}
{Barnes}, J.~W., \& {O'Brien}, D.~P. 2002,
  \href{http://dx.doi.org/10.1086/341477}{\JournalTitle{\apj}, 575, 1087}

\bibitem[{Barstow {et~al.}(2017)}]{Barstow2017}
Barstow, J.~K., {et~al.} 2017, \JournalTitle{ApJ}, 834

\bibitem[{{Ben-Jaffel} \& {Ballester}(2014)}]{bj2014}
{Ben-Jaffel}, L., \& {Ballester}, G.~E. 2014,
  \href{http://dx.doi.org/10.1088/2041-8205/785/2/L30}{\JournalTitle{\apjl},
  785, L30}

\bibitem[{{Ben-Jaffel} \& {Sona Hosseini}(2010)}]{bj2010}
{Ben-Jaffel}, L., \& {Sona Hosseini}, S. 2010,
  \href{http://dx.doi.org/10.1088/0004-637X/709/2/1284}{\JournalTitle{\apj},
  709, 1284}

\bibitem[{{B{\'e}zard} {et~al.}(2002){B{\'e}zard}, {Lellouch}, {Strobel},
  {Maillard}, \& {Drossart}}]{bezard2002}
{B{\'e}zard}, B., {Lellouch}, E., {Strobel}, D., {Maillard}, J.-P., \&
  {Drossart}, P. 2002,
  \href{http://dx.doi.org/10.1006/icar.2002.6917}{\JournalTitle{\icarus}, 159,
  95}

\bibitem[{{Bower} {et~al.}(2019){Bower}, {Kitzmann}, {Wolf}, {Sanan}, {Dorn},
  \& {Oza}}]{Bower19}
{Bower}, D.~J., {Kitzmann}, D., {Wolf}, A.~S., {et~al.} 2019,
  \JournalTitle{arXiv e-prints}, arXiv:1904.08300

\bibitem[{{Boyajian} {et~al.}(2016){Boyajian}, {LaCourse}, {Rappaport},
  {Fabrycky}, {Fischer}, {Gandolfi}, {Kennedy}, {Korhonen}, {Liu}, {Moor},
  {Olah}, {Vida}, {Wyatt}, {Best}, {Brewer}, {Ciesla}, {Cs{\'a}k}, {Deeg},
  {Dupuy}, {Handler}, {Heng}, {Howell}, {Ishikawa}, {Kov{\'a}cs}, {Kozakis},
  {Kriskovics}, {Lehtinen}, {Lintott}, {Lynn}, {Nespral}, {Nikbakhsh},
  {Schawinski}, {Schmitt}, {Smith}, {Szabo}, {Szabo}, {Viuho}, {Wang},
  {Weiksnar}, {Bosch}, {Connors}, {Goodman}, {Green}, {Hoekstra}, {Jebson},
  {Jek}, {Omohundro}, {Schwengeler}, \& {Szewczyk}}]{boyajian16}
{Boyajian}, T.~S., {LaCourse}, D.~M., {Rappaport}, S.~A., {et~al.} 2016,
  \href{http://dx.doi.org/10.1093/mnras/stw218}{\JournalTitle{\mnras}, 457,
  3988}

\bibitem[{{Brown}(2001{\natexlab{a}})}]{brown2001}
{Brown}, M.~E. 2001{\natexlab{a}},
  \href{http://dx.doi.org/10.1006/icar.2001.6612}{\JournalTitle{\icarus}, 151,
  190}

\bibitem[{{Brown}(2001{\natexlab{b}})}]{brown01}
---. 2001{\natexlab{b}},
  \href{http://dx.doi.org/10.1006/icar.2001.6612}{\JournalTitle{\icarus}, 151,
  190}

\bibitem[{{Brown} \& {Hill}(1996)}]{brown96}
{Brown}, M.~E., \& {Hill}, R.~E. 1996,
  \href{http://dx.doi.org/10.1038/380229a0}{\JournalTitle{\nat}, 380, 229}

\bibitem[{{Brown}(1974)}]{brown74}
{Brown}, R.~A. 1974, in IAU Symposium, Vol.~65, Exploration of the Planetary
  System, ed. A.~{Woszczyk} \& C.~{Iwaniszewska}, 527

\bibitem[{{Brown} \& {Chaffee}(1974)}]{brownchaffee74}
{Brown}, R.~A., \& {Chaffee}, Jr., F.~H. 1974,
  \href{http://dx.doi.org/10.1086/181413}{\JournalTitle{\apjl}, 187, L125}

\bibitem[{{Burger} {et~al.}(2001){Burger}, {Schneider}, {de Pater}, {Brown},
  {Bouchez}, {Trafton}, {Sheffer}, {Barker}, \& {Mallama}}]{burger01}
{Burger}, M.~H., {Schneider}, N.~M., {de Pater}, I., {et~al.} 2001,
  \href{http://dx.doi.org/10.1086/323944}{\JournalTitle{\apj}, 563, 1063}

\bibitem[{{Burns} {et~al.}(1999){Burns}, {Showalter}, {Hamilton}, {Nicholson},
  {de Pater}, {Ockert-Bell}, \& {Thomas}}]{burns99}
{Burns}, J.~A., {Showalter}, M.~R., {Hamilton}, D.~P., {et~al.} 1999,
  \href{http://dx.doi.org/10.1126/science.284.5417.1146}{\JournalTitle{Science},
  284, 1146}

\bibitem[{Casasayas-Barris {et~al.}(2017)}]{Casasayas-Barris2017}
Casasayas-Barris, N., {et~al.} 2017, \JournalTitle{A\&A}, 608

\bibitem[{{Cassidy} {et~al.}(2009){Cassidy}, {Mendez}, {Arras}, {Johnson}, \&
  {Skrutskie}}]{cassidy09}
{Cassidy}, T.~A., {Mendez}, R., {Arras}, P., {Johnson}, R.~E., \& {Skrutskie},
  M.~F. 2009,
  \href{http://dx.doi.org/10.1088/0004-637X/704/2/1341}{\JournalTitle{\apj},
  704, 1341}

\bibitem[{{Chamberlain}(1961)}]{chamberlain61}
{Chamberlain}, J.~W. 1961, {Physics of the aurora and airglow}

\bibitem[{{Chandrasekhar}(1969)}]{chandrasekhar69}
{Chandrasekhar}, S. 1969, {Ellipsoidal figures of equilibrium}

\bibitem[{{Charbonneau} {et~al.}(2002){Charbonneau}, {Brown}, {Noyes}, \&
  {Gilliland}}]{charbonneau02}
{Charbonneau}, D., {Brown}, T.~M., {Noyes}, R.~W., \& {Gilliland}, R.~L. 2002,
  \href{http://dx.doi.org/10.1086/338770}{\JournalTitle{\apj}, 568, 377}

\bibitem[{Chen {et~al.}(2017)}]{Chen2017}
Chen, G., {et~al.} 2017, \JournalTitle{A\&A}, 600

\bibitem[{{Christensen} {et~al.}(2009){Christensen}, {Holzwarth}, \&
  {Reiners}}]{christensen09}
{Christensen}, U.~R., {Holzwarth}, V., \& {Reiners}, A. 2009,
  \href{http://dx.doi.org/10.1038/nature07626}{\JournalTitle{\nat}, 457, 167}

\bibitem[{Col\'{o}n {et~al.}(2012)}]{Colon2012}
Col\'{o}n, K.~D., {et~al.} 2012, \JournalTitle{MNRAS}, 419

\bibitem[{{Crovisier}(1996)}]{crovisier96}
{Crovisier}, J. 1996, in IAU Colloq. 156: The Collision of Comet Shoemaker-Levy
  9 and Jupiter, ed. K.~S. {Noll}, H.~A. {Weaver}, \& P.~D. {Feldman}, 31

\bibitem[{{Cubillos} {et~al.}(2017){Cubillos}, {Fossati}, {Erkaev}, {Malik},
  {Tokano}, {Lendl}, {Johnstone}, {Lammer}, \& {Wyttenbach}}]{cubillos17}
{Cubillos}, P.~E., {Fossati}, L., {Erkaev}, N.~V., {et~al.} 2017,
  \href{http://dx.doi.org/10.3847/1538-4357/aa9019}{\JournalTitle{\apj}, 849,
  145}

\bibitem[{{de Kleer} \& {de Pater}(2016)}]{dekleer16}
{de Kleer}, K., \& {de Pater}, I. 2016,
  \href{http://dx.doi.org/10.1016/j.icarus.2016.06.019}{\JournalTitle{\icarus},
  280, 378}

\bibitem[{{de Pater} {et~al.}(2017){de Pater}, {de Kleer}, {Davies}, \&
  {{\'A}d{\'a}mkovics}}]{imke17}
{de Pater}, I., {de Kleer}, K., {Davies}, A.~G., \& {{\'A}d{\'a}mkovics}, M.
  2017,
  \href{http://dx.doi.org/10.1016/j.icarus.2017.03.016}{\JournalTitle{\icarus},
  297, 265}

\bibitem[{{Domingos} {et~al.}(2006){Domingos}, {Winter}, \&
  {Yokoyama}}]{domingos06}
{Domingos}, R.~C., {Winter}, O.~C., \& {Yokoyama}, T. 2006,
  \href{http://dx.doi.org/10.1111/j.1365-2966.2006.11104.x}{\JournalTitle{\mnras},
  373, 1227}

\bibitem[{{Draine}(2011)}]{draine2011}
{Draine}, B.~T. 2011, {Physics of the Interstellar and Intergalactic Medium}

\bibitem[{{Fegley} \& {Cameron}(1987)}]{fc87}
{Fegley}, B., \& {Cameron}, A.~G.~W. 1987,
  \href{http://dx.doi.org/10.1016/0012-821X(87)90196-8}{\JournalTitle{Earth and
  Planetary Science Letters}, 82, 207}

\bibitem[{{Fegley} \& {Zolotov}(2000)}]{fegleyzolotov2000}
{Fegley}, B., \& {Zolotov}, M.~Y. 2000,
  \href{http://dx.doi.org/10.1006/icar.2000.6490}{\JournalTitle{\icarus}, 148,
  193}

\bibitem[{{Fischer} {et~al.}(2016){Fischer}, {Knutson}, {Sing}, {Henry},
  {Williamson}, {Fortney}, {Burrows}, {Kataria}, {Nikolov}, {Showman},
  {Ballester}, {D{\'e}sert}, {Aigrain}, {Deming}, {Lecavelier des Etangs}, \&
  {Vidal-Madjar}}]{Fischer2016}
{Fischer}, P.~D., {Knutson}, H.~A., {Sing}, D.~K., {et~al.} 2016,
  \href{http://dx.doi.org/10.3847/0004-637X/827/1/19}{\JournalTitle{\apj}, 827,
  19}

\bibitem[{{Fisher} \& {Heng}(2019)}]{fisher_heng2019}
{Fisher}, C., \& {Heng}, K. 2019,
  \href{http://dx.doi.org/10.3847/1538-4357/ab29e8}{\JournalTitle{The
  Astrophysical Journal}, 881, 25}

\bibitem[{{Fossati} {et~al.}(2010){Fossati}, {Haswell}, {Froning}, {Hebb},
  {Holmes}, {Kolb}, {Helling}, {Carter}, {Wheatley}, {Collier Cameron},
  {Loeillet}, {Pollacco}, {Street}, {Stempels}, {Simpson}, {Udry}, {Joshi},
  {West}, {Skillen}, \& {Wilson}}]{mg2}
{Fossati}, L., {Haswell}, C.~A., {Froning}, C.~S., {et~al.} 2010,
  \href{http://dx.doi.org/10.1088/2041-8205/714/2/L222}{\JournalTitle{\apj},
  714, L222}

\bibitem[{{Fulle} {et~al.}(2013){Fulle}, {Molaro}, {Buzzi}, \&
  {Valisa}}]{fulle2013}
{Fulle}, M., {Molaro}, P., {Buzzi}, L., \& {Valisa}, P. 2013,
  \href{http://dx.doi.org/10.1088/2041-8205/771/2/L21}{\JournalTitle{\apjl},
  771, L21}

\bibitem[{{Fulton} \& {Petigura}(2018)}]{fulton18}
{Fulton}, B.~J., \& {Petigura}, E.~A. 2018,
  \href{http://dx.doi.org/10.3847/1538-3881/aae828}{\JournalTitle{\aj}, 156,
  264}

\bibitem[{{Gardner} {et~al.}(2014){Gardner}, {Liu}, {Marsh}, {Feng}, \&
  {Plane}}]{gardner14}
{Gardner}, C.~S., {Liu}, A.~Z., {Marsh}, D.~R., {Feng}, W., \& {Plane},
  J.~M.~C. 2014,
  \href{http://dx.doi.org/10.1002/2014JA020383}{\JournalTitle{Journal of
  Geophysical Research (Space Physics)}, 119, 7870}

\bibitem[{{Gibson} {et~al.}(2019){Gibson}, {de Mooij}, {Evans}, {Merritt},
  {Nikolov}, {Sing}, \& {Watson}}]{gibson19}
{Gibson}, N.~P., {de Mooij}, E.~J.~W., {Evans}, T.~M., {et~al.} 2019,
  \href{http://dx.doi.org/10.1093/mnras/sty2722}{\JournalTitle{\mnras}, 482,
  606}

\bibitem[{{Goldreich} \& {Nicholson}(1977)}]{goldreich77}
{Goldreich}, P., \& {Nicholson}, P.~D. 1977,
  \href{http://dx.doi.org/10.1016/0019-1035(77)90163-4}{\JournalTitle{\icarus},
  30, 301}

\bibitem[{{Grie{\ss}meier} {et~al.}(2007){Grie{\ss}meier}, {Zarka}, \&
  {Spreeuw}}]{Griessmeier07}
{Grie{\ss}meier}, J.-M., {Zarka}, P., \& {Spreeuw}, H. 2007,
  \href{http://dx.doi.org/10.1051/0004-6361:20077397}{\JournalTitle{\aap}, 475,
  359}

\bibitem[{{Haff} {et~al.}(1981){Haff}, {Watson}, \& {Yung}}]{haff81}
{Haff}, P.~K., {Watson}, C.~C., \& {Yung}, Y.~L. 1981,
  \href{http://dx.doi.org/10.1029/JA086iA08p06933}{\JournalTitle{\jgr}, 86,
  6933}

\bibitem[{Hanson \& Donaldson(1967)}]{hansondonaldson67}
Hanson, W.~B., \& Donaldson, J.~S. 1967,
  \href{http://dx.doi.org/10.1029/JZ072i021p05513}{\JournalTitle{Journal of
  Geophysical Research}, 72, 5513}

\bibitem[{{Heng}(2016)}]{heng16}
{Heng}, K. 2016,
  \href{http://dx.doi.org/10.3847/2041-8205/826/1/L16}{\JournalTitle{\apjl},
  826, L16}

\bibitem[{{Heng} {et~al.}(2015){Heng}, {Wyttenbach}, {Lavie}, {Sing},
  {Ehrenreich}, \& {Lovis}}]{heng15}
{Heng}, K., {Wyttenbach}, A., {Lavie}, B., {et~al.} 2015,
  \href{http://dx.doi.org/10.1088/2041-8205/803/1/L9}{\JournalTitle{\apjl},
  803, L9}

\bibitem[{{Hoeijmakers} {et~al.}(2018){Hoeijmakers}, {Ehrenreich}, {Heng},
  {Kitzmann}, {Grimm}, {Allart}, {Deitrick}, {Wyttenbach}, {Oreshenko}, {Pino},
  {Rimmer}, {Molinari}, \& {Di Fabrizio}}]{jens2018}
{Hoeijmakers}, H.~J., {Ehrenreich}, D., {Heng}, K., {et~al.} 2018,
  \href{http://dx.doi.org/10.1038/s41586-018-0401-y}{\JournalTitle{\nat}, 560,
  453}

\bibitem[{{Huang} {et~al.}(2017){Huang}, {Arras}, {Christie}, \&
  {Li}}]{huang17}
{Huang}, C., {Arras}, P., {Christie}, D., \& {Li}, Z.-Y. 2017,
  \href{http://dx.doi.org/10.3847/1538-4357/aa9b32}{\JournalTitle{\apj}, 851,
  150}

\bibitem[{{Huebner} \& {Mukherjee}(2015)}]{huebnermujherjee}
{Huebner}, W.~F., \& {Mukherjee}, J. 2015,
  \href{http://dx.doi.org/10.1016/j.pss.2014.11.022}{\JournalTitle{\planss},
  106, 11}

\bibitem[{{Huitson} {et~al.}(2012){Huitson}, {Sing}, {Vidal-Madjar},
  {Ballester}, {Lecavelier des Etangs}, {D{\'e}sert}, \& {Pont}}]{huitson2012}
{Huitson}, C.~M., {Sing}, D.~K., {Vidal-Madjar}, A., {et~al.} 2012,
  \href{http://dx.doi.org/10.1111/j.1365-2966.2012.20805.x}{\JournalTitle{\mnras},
  422, 2477}

\bibitem[{{Hunten}(1967)}]{naearth}
{Hunten}, D.~M. 1967,
  \href{http://dx.doi.org/10.1007/BF00173704}{\JournalTitle{\ssr}, 6, 493}

\bibitem[{Hunten \& Wallace(1967)}]{huntenwallace67}
Hunten, D.~M., \& Wallace, L. 1967,
  \href{http://dx.doi.org/10.1029/JZ072i001p00069}{\JournalTitle{Journal of
  Geophysical Research}, 72, 69}

\bibitem[{{Ingersoll}(1989)}]{ingersoll89}
{Ingersoll}, A.~P. 1989,
  \href{http://dx.doi.org/10.1016/0019-1035(89)90055-9}{\JournalTitle{\icarus},
  81, 298}

\bibitem[{{Jin} \& {Mordasini}(2018)}]{jinmordasini18}
{Jin}, S., \& {Mordasini}, C. 2018,
  \href{http://dx.doi.org/10.3847/1538-4357/aa9f1e}{\JournalTitle{\apj}, 853,
  163}

\bibitem[{{Jin} {et~al.}(2014){Jin}, {Mordasini}, {Parmentier}, {van Boekel},
  {Henning}, \& {Ji}}]{jinmordasini14}
{Jin}, S., {Mordasini}, C., {Parmentier}, V., {et~al.} 2014,
  \href{http://dx.doi.org/10.1088/0004-637X/795/1/65}{\JournalTitle{\apj}, 795,
  65}

\bibitem[{{Johnson}(1990)}]{bob90book}
{Johnson}, R.~E. 1990, {Energetic Charged-Particle Interactions with
  Atmospheres and Surfaces}, 84

\bibitem[{{Johnson}(2004)}]{johnson04}
---. 2004, \href{http://dx.doi.org/10.1086/422912}{\JournalTitle{\apjl}, 609,
  L99}

\bibitem[{{Johnson} \& {Huggins}(2006)}]{johnsonhuggins06}
{Johnson}, R.~E., \& {Huggins}, P.~J. 2006,
  \href{http://dx.doi.org/10.1086/506183}{\JournalTitle{\pasp}, 118, 1136}

\bibitem[{{Johnson} {et~al.}(2002){Johnson}, {Leblanc}, {Yakshinskiy}, \&
  {Madey}}]{johnson02}
{Johnson}, R.~E., {Leblanc}, F., {Yakshinskiy}, B.~V., \& {Madey}, T.~E. 2002,
  \href{http://dx.doi.org/10.1006/icar.2001.6763}{\JournalTitle{\icarus}, 156,
  136}

\bibitem[{{Johnson} {et~al.}(2015){Johnson}, {Oza}, {Young}, {Volkov}, \&
  {Schmidt}}]{johnson15}
{Johnson}, R.~E., {Oza}, A., {Young}, L.~A., {Volkov}, A.~N., \& {Schmidt}, C.
  2015,
  \href{http://dx.doi.org/10.1088/0004-637X/809/1/43}{\JournalTitle{\apj}, 809,
  43}

\bibitem[{{Johnson} {et~al.}(2006{\natexlab{a}}){Johnson}, {Smith}, {Tucker},
  {Liu}, {Burger}, {Sittler}, \& {Tokar}}]{johnson06b}
{Johnson}, R.~E., {Smith}, H.~T., {Tucker}, O.~J., {et~al.} 2006{\natexlab{a}},
  \href{http://dx.doi.org/10.1086/505750}{\JournalTitle{\apjl}, 644, L137}

\bibitem[{{Johnson} {et~al.}(2013){Johnson}, {Volkov}, \& {Erwin}}]{johnson13}
{Johnson}, R.~E., {Volkov}, A.~N., \& {Erwin}, J.~T. 2013,
  \href{http://dx.doi.org/10.1088/2041-8205/768/1/L4}{\JournalTitle{\apjl},
  768, L4}

\bibitem[{{Johnson} {et~al.}(2006{\natexlab{b}}){Johnson}, {Luhmann}, {Tokar},
  {Bouhram}, {Berthelier}, {Sittler}, {Cooper}, {Hill}, {Smith}, {Michael},
  {Liu}, {Crary}, \& {Young}}]{johnson06a}
{Johnson}, R.~E., {Luhmann}, J.~G., {Tokar}, R.~L., {et~al.}
  2006{\natexlab{b}},
  \href{http://dx.doi.org/10.1016/j.icarus.2005.08.021}{\JournalTitle{\icarus},
  180, 393}

\bibitem[{{Kao} {et~al.}(2018){Kao}, {Hallinan}, {Pineda}, {Stevenson}, \&
  {Burgasser}}]{kao18}
{Kao}, M.~M., {Hallinan}, G., {Pineda}, J.~S., {Stevenson}, D., \& {Burgasser},
  A. 2018,
  \href{http://dx.doi.org/10.3847/1538-4365/aac2d5}{\JournalTitle{\apjs}, 237,
  25}

\bibitem[{{Khalafinejad} {et~al.}(2017){Khalafinejad}, {von Essen},
  {Hoeijmakers}, {Zhou}, {Klocov{\'a}}, {Schmitt}, {Dreizler}, {Lopez-Morales},
  {Husser}, {Schmidt}, \& {Collet}}]{khalafinejad17}
{Khalafinejad}, S., {von Essen}, C., {Hoeijmakers}, H.~J., {et~al.} 2017,
  \href{http://dx.doi.org/10.1051/0004-6361/201629473}{\JournalTitle{\aap},
  598, A131}

\bibitem[{Khalafinejad {et~al.}(2018)}]{Khalafinejad2018}
Khalafinejad, S., {et~al.} 2018, apJ

\bibitem[{{Killen} {et~al.}(2001){Killen}, {Potter}, {Reiff}, {Sarantos},
  {Jackson}, {Hick}, \& {Giles}}]{killen2001}
{Killen}, R.~M., {Potter}, A.~E., {Reiff}, P., {et~al.} 2001,
  \href{http://dx.doi.org/10.1029/2000JE001401}{\JournalTitle{\jgr}, 106,
  20509}

\bibitem[{{Kipping}(2009)}]{kippingexomoons09}
{Kipping}, D.~M. 2009,
  \href{http://dx.doi.org/10.1111/j.1365-2966.2008.13999.x}{\JournalTitle{\mnras},
  392, 181}

\bibitem[{{Kislyakova} {et~al.}(2014){Kislyakova}, {Holmstr{\"o}m}, {Lammer},
  {Odert}, \& {Khodachenko}}]{kislyakova14}
{Kislyakova}, K.~G., {Holmstr{\"o}m}, M., {Lammer}, H., {Odert}, P., \&
  {Khodachenko}, M.~L. 2014,
  \href{http://dx.doi.org/10.1126/science.1257829}{\JournalTitle{Science}, 346,
  981}

\bibitem[{{Kislyakova} {et~al.}(2016){Kislyakova}, {Pilat-Lohinger}, {Funk},
  {Lammer}, {Fossati}, {Eggl}, {Schwarz}, {Boudjada}, \&
  {Erkaev}}]{kislyakova16}
{Kislyakova}, K.~G., {Pilat-Lohinger}, E., {Funk}, B., {et~al.} 2016,
  \href{http://dx.doi.org/10.1093/mnras/stw1110}{\JournalTitle{\mnras}, 461,
  988}

\bibitem[{{Kreidberg} {et~al.}(2019){Kreidberg}, {Luger}, \&
  {Bedell}}]{kreidberg2019}
{Kreidberg}, L., {Luger}, R., \& {Bedell}, M. 2019,
  \href{http://dx.doi.org/10.3847/2041-8213/ab20c8}{\JournalTitle{\apj}, 877,
  L15}

\bibitem[{{Kreidberg} {et~al.}(2014){Kreidberg}, {Bean}, {D{\'e}sert}, {Line},
  {Fortney}, {Madhusudhan}, {Stevenson}, {Showman}, {Charbonneau},
  {McCullough}, {Seager}, {Burrows}, {Henry}, {Williamson}, {Kataria}, \&
  {Homeier}}]{kreidberg2014b}
{Kreidberg}, L., {Bean}, J.~L., {D{\'e}sert}, J.-M., {et~al.} 2014,
  \href{http://dx.doi.org/10.1088/2041-8205/793/2/L27}{\JournalTitle{\apjl},
  793, L27}

\bibitem[{{K{\"u}ppers} \& {Schneider}(2000)}]{kuppers2000}
{K{\"u}ppers}, M., \& {Schneider}, N.~M. 2000,
  \href{http://dx.doi.org/10.1029/1999GL010718}{\JournalTitle{\grl}, 27, 513}

\bibitem[{{Lainey} \& {Tobie}(2005)}]{laineytobie05}
{Lainey}, V., \& {Tobie}, G. 2005,
  \href{http://dx.doi.org/10.1016/j.icarus.2005.07.017}{\JournalTitle{\icarus},
  179, 485}

\bibitem[{{Lammer} {et~al.}(2009){Lammer}, {Odert}, {Leitzinger},
  {Khodachenko}, {Panchenko}, {Kulikov}, {Zhang}, {Lichtenegger}, {Erkaev}, \&
  {Wuchterl}}]{lammer09}
{Lammer}, H., {Odert}, P., {Leitzinger}, M., {et~al.} 2009,
  \href{http://dx.doi.org/10.1051/0004-6361/200911922}{\JournalTitle{\aap},
  506, 399}

\bibitem[{{Langland-Shula} {et~al.}(2009){Langland-Shula}, {Vogt},
  {Charbonneau}, {Butler}, \& {Marcy}}]{Langland-Shula09}
{Langland-Shula}, L.~E., {Vogt}, S.~S., {Charbonneau}, D., {Butler}, P., \&
  {Marcy}, G. 2009,
  \href{http://dx.doi.org/10.1088/0004-637X/696/2/1355}{\JournalTitle{\apj},
  696, 1355}

\bibitem[{{Leblanc} \& {Johnson}(2010)}]{leblancjohnson10}
{Leblanc}, F., \& {Johnson}, R.~E. 2010,
  \href{http://dx.doi.org/10.1016/j.icarus.2010.04.020}{\JournalTitle{\icarus},
  209, 280}

\bibitem[{{Leblanc} {et~al.}(2002){Leblanc}, {Johnson}, \& {Brown}}]{leblanc02}
{Leblanc}, F., {Johnson}, R.~E., \& {Brown}, M.~E. 2002,
  \href{http://dx.doi.org/10.1006/icar.2002.6934}{\JournalTitle{\icarus}, 159,
  132}

\bibitem[{{Leblanc} {et~al.}(2005){Leblanc}, {Potter}, {Killen}, \&
  {Johnson}}]{leblanc05}
{Leblanc}, F., {Potter}, A.~E., {Killen}, R.~M., \& {Johnson}, R.~E. 2005,
  \href{http://dx.doi.org/10.1016/j.icarus.2005.03.027}{\JournalTitle{\icarus},
  178, 367}

\bibitem[{{Leblanc} {et~al.}(2008){Leblanc}, {Doressoundiram}, {Schneider},
  {Mangano}, {L{\'o}pez Ariste}, {Lemen}, {Gelly}, {Barbieri}, \&
  {Cremonese}}]{leblanc08}
{Leblanc}, F., {Doressoundiram}, A., {Schneider}, N., {et~al.} 2008,
  \href{http://dx.doi.org/10.1029/2008GL035322}{\JournalTitle{\grl}, 35,
  L18204}

\bibitem[{{Lellouch}(1996)}]{lellouch96}
{Lellouch}, E. 1996, in IAU Colloq. 156: The Collision of Comet Shoemaker-Levy
  9 and Jupiter, ed. K.~S. {Noll}, H.~A. {Weaver}, \& P.~D. {Feldman}, 213

\bibitem[{{Lellouch} {et~al.}(2015){Lellouch}, {Ali-Dib}, {Jessup}, {Smette},
  {K{\"a}ufl}, \& {Marchis}}]{lellouch15}
{Lellouch}, E., {Ali-Dib}, M., {Jessup}, K.-L., {et~al.} 2015,
  \href{http://dx.doi.org/10.1016/j.icarus.2015.02.018}{\JournalTitle{\icarus},
  253, 99}

\bibitem[{{Lellouch} {et~al.}(1990){Lellouch}, {Belton}, {de Pater}, {Gulkis},
  \& {Encrenaz}}]{lellouch90}
{Lellouch}, E., {Belton}, M., {de Pater}, I., {Gulkis}, S., \& {Encrenaz}, T.
  1990, \href{http://dx.doi.org/10.1038/346639a0}{\JournalTitle{\nat}, 346,
  639}

\bibitem[{{Lellouch} {et~al.}(1992){Lellouch}, {Belton}, {de Pater}, {Paubert},
  {Gulkis}, \& {Encrenaz}}]{lellouch92}
{Lellouch}, E., {Belton}, M., {de Pater}, I., {et~al.} 1992,
  \href{http://dx.doi.org/10.1016/0019-1035(92)90095-O}{\JournalTitle{\icarus},
  98, 271}

\bibitem[{{Lellouch} {et~al.}(2003){Lellouch}, {Paubert}, {Moses}, {Schneider},
  \& {Strobel}}]{lellouch03}
{Lellouch}, E., {Paubert}, G., {Moses}, J.~I., {Schneider}, N.~M., \&
  {Strobel}, D.~F. 2003,
  \href{http://dx.doi.org/10.1038/nature01292}{\JournalTitle{\nat}, 421, 45}

\bibitem[{{Lellouch} {et~al.}(1996){Lellouch}, {Strobel}, {Belton}, {Summers},
  {Paubert}, \& {Moreno}}]{lellouch96SO}
{Lellouch}, E., {Strobel}, D.~F., {Belton}, M.~J.~S., {et~al.} 1996,
  \href{http://dx.doi.org/10.1086/309956}{\JournalTitle{\apjl}, 459, L107}

\bibitem[{{Lellouch} {et~al.}(1997){Lellouch}, {B{\'e}zard}, {Moreno},
  {Bockel{\'e}e-Morvan}, {Colom}, {Crovisier}, {Festou}, {Gautier}, {Marten},
  \& {Paubert}}]{lellouch97}
{Lellouch}, E., {B{\'e}zard}, B., {Moreno}, R., {et~al.} 1997,
  \href{http://dx.doi.org/10.1016/S0032-0633(97)00043-3}{\JournalTitle{\planss},
  45, 1203}

\bibitem[{{Lellouch} {et~al.}(2002){Lellouch}, {B{\'e}zard}, {Moses}, {Davis},
  {Drossart}, {Feuchtgruber}, {Bergin}, {Moreno}, \& {Encrenaz}}]{lellouch02}
{Lellouch}, E., {B{\'e}zard}, B., {Moses}, J.~I., {et~al.} 2002,
  \href{http://dx.doi.org/10.1006/icar.2002.6929}{\JournalTitle{\icarus}, 159,
  112}

\bibitem[{{Lide}(1994)}]{lide94}
{Lide}, D.~R. 1994, {CRC Handbook of chemistry and physics. A ready-reference
  book of chemical and physical data}

\bibitem[{{Lodders}(2010)}]{lodders2010}
{Lodders}, K. 2010,
  \href{http://dx.doi.org/10.1007/978-3-642-10352-0_8}{\JournalTitle{Astrophysics
  and Space Science Proceedings}, 16, 379}

\bibitem[{{Louden} \& {Wheatley}(2015)}]{loudenwheatley15}
{Louden}, T., \& {Wheatley}, P.~J. 2015,
  \href{http://dx.doi.org/10.1088/2041-8205/814/2/L24}{\JournalTitle{\apjl},
  814, L24}

\bibitem[{{Marchis} {et~al.}(2005){Marchis}, {Le Mignant}, {Chaffee}, {Davies},
  {Kwok}, {Prang{\'e}}, {de Pater}, {Amico}, {Campbell}, {Fusco}, {Goodrich},
  \& {Conrad}}]{marchis05}
{Marchis}, F., {Le Mignant}, D., {Chaffee}, F.~H., {et~al.} 2005,
  \href{http://dx.doi.org/10.1016/j.icarus.2004.12.014}{\JournalTitle{\icarus},
  176, 96}

\bibitem[{{Matsakos} {et~al.}(2015){Matsakos}, {Uribe}, \&
  {K{\"o}nigl}}]{matsakos15}
{Matsakos}, T., {Uribe}, A., \& {K{\"o}nigl}, A. 2015,
  \href{http://dx.doi.org/10.1051/0004-6361/201425593}{\JournalTitle{\aap},
  578, A6}

\bibitem[{{McDoniel} {et~al.}(2017){McDoniel}, {Goldstein}, {Varghese}, \&
  {Trafton}}]{mcdoniel17}
{McDoniel}, W.~J., {Goldstein}, D.~B., {Varghese}, P.~L., \& {Trafton}, L.~M.
  2017,
  \href{http://dx.doi.org/10.1016/j.icarus.2017.04.021}{\JournalTitle{\icarus},
  294, 81}

\bibitem[{{Mendillo} {et~al.}(1990){Mendillo}, {Baumgardner}, {Flynn}, \&
  {Hughes}}]{mendillo90}
{Mendillo}, M., {Baumgardner}, J., {Flynn}, B., \& {Hughes}, W.~J. 1990,
  \href{http://dx.doi.org/10.1038/348312a0}{\JournalTitle{\nat}, 348, 312}

\bibitem[{{Monta{\~n}{\'e}s-Rodr{\'{\i}}guez}
  {et~al.}(2015){Monta{\~n}{\'e}s-Rodr{\'{\i}}guez}, {Gonz{\'a}lez-Merino},
  {Pall{\'e}}, {L{\'o}pez-Puertas}, \& {Garc{\'{\i}}a-Melendo}}]{montanes}
{Monta{\~n}{\'e}s-Rodr{\'{\i}}guez}, P., {Gonz{\'a}lez-Merino}, B.,
  {Pall{\'e}}, E., {L{\'o}pez-Puertas}, M., \& {Garc{\'{\i}}a-Melendo}, E.
  2015,
  \href{http://dx.doi.org/10.1088/2041-8205/801/1/L8}{\JournalTitle{\apjl},
  801, L8}

\bibitem[{{Morabito} {et~al.}(1979){Morabito}, {Synnott}, {Kupferman}, \&
  {Collins}}]{morabito79}
{Morabito}, L.~A., {Synnott}, S.~P., {Kupferman}, P.~N., \& {Collins}, S.~A.
  1979,
  \href{http://dx.doi.org/10.1126/science.204.4396.972}{\JournalTitle{Science},
  204, 972}

\bibitem[{{Mordasini} {et~al.}(2012){Mordasini}, {Alibert}, {Benz}, {Klahr}, \&
  {Henning}}]{mordasini2012metals}
{Mordasini}, C., {Alibert}, Y., {Benz}, W., {Klahr}, H., \& {Henning}, T. 2012,
  \href{http://dx.doi.org/10.1051/0004-6361/201117350}{\JournalTitle{\aap},
  541, A97}

\bibitem[{{Moreno} {et~al.}(2003){Moreno}, {Marten}, {Matthews}, \&
  {Biraud}}]{moreno03}
{Moreno}, R., {Marten}, A., {Matthews}, H.~E., \& {Biraud}, Y. 2003,
  \href{http://dx.doi.org/10.1016/S0032-0633(03)00072-2}{\JournalTitle{\planss},
  51, 591}

\bibitem[{{Moses} \& {Poppe}(2017)}]{mosespoppe2017}
{Moses}, J.~I., \& {Poppe}, A.~R. 2017,
  \href{http://dx.doi.org/10.1016/j.icarus.2017.06.002}{\JournalTitle{\icarus},
  297, 33}

\bibitem[{{Moullet} {et~al.}(2015){Moullet}, {Lellouch}, {Gurwell}, {Moreno},
  {Black}, \& {Butler}}]{moullet15}
{Moullet}, A., {Lellouch}, E., {Gurwell}, M., {et~al.} 2015, in AAS/Division
  for Planetary Sciences Meeting Abstracts, Vol.~47, AAS/Division for Planetary
  Sciences Meeting Abstracts \#47, 311.31

\bibitem[{{Moullet} {et~al.}(2013){Moullet}, {Lellouch}, {Moreno}, {Gurwell},
  {Black}, \& {Butler}}]{moullet13}
{Moullet}, A., {Lellouch}, E., {Moreno}, R., {et~al.} 2013,
  \href{http://dx.doi.org/10.1088/0004-637X/776/1/32}{\JournalTitle{\apj}, 776,
  32}

\bibitem[{{Murray-Clay} {et~al.}(2009){Murray-Clay}, {Chiang}, \&
  {Murray}}]{mc09}
{Murray-Clay}, R.~A., {Chiang}, E.~I., \& {Murray}, N. 2009,
  \href{http://dx.doi.org/10.1088/0004-637X/693/1/23}{\JournalTitle{\apj}, 693,
  23}

\bibitem[{{Nikolov} {et~al.}(2014){Nikolov}, {Sing}, {Pont}, {Burrows},
  {Fortney}, {Ballester}, {Evans}, {Huitson}, {Wakeford}, {Wilson}, {Aigrain},
  {Deming}, {Gibson}, {Henry}, {Knutson}, {Lecavelier des Etangs}, {Showman},
  {Vidal-Madjar}, \& {Zahnle}}]{nikolov14}
{Nikolov}, N., {Sing}, D.~K., {Pont}, F., {et~al.} 2014,
  \href{http://dx.doi.org/10.1093/mnras/stt1859}{\JournalTitle{\mnras}, 437,
  46}

\bibitem[{Nikolov {et~al.}(2016)}]{Nikolov2016}
Nikolov, N., {et~al.} 2016, \JournalTitle{ApJ}, 832

\bibitem[{{Nikolov} {et~al.}(2018){Nikolov}, {Sing}, {Fortney}, {Goyal},
  {Drummond}, {Evans}, {Gibson}, {De Mooij}, {Rustamkulov}, {Wakeford},
  {Smalley}, {Burgasser}, {Hellier}, {Helling}, {Mayne}, {Madhusudhan},
  {Kataria}, {Baines}, {Carter}, {Ballester}, {Barstow}, {McCleery}, \&
  {Spake}}]{nikolov18}
{Nikolov}, N., {Sing}, D.~K., {Fortney}, J.~J., {et~al.} 2018,
  \href{http://dx.doi.org/10.1038/s41586-018-0101-7}{\JournalTitle{\nat}, 557,
  526}

\bibitem[{{Noack} {et~al.}(2017){Noack}, {Rivoldini}, \& {Van
  Hoolst}}]{noack17}
{Noack}, L., {Rivoldini}, A., \& {Van Hoolst}, T. 2017,
  \href{http://dx.doi.org/10.1016/j.pepi.2017.05.010}{\JournalTitle{Physics of
  the Earth and Planetary Interiors}, 269, 40}

\bibitem[{{Noll} {et~al.}(1995){Noll}, {Geballe}, \& {Knacke}}]{noll95}
{Noll}, K.~S., {Geballe}, T.~R., \& {Knacke}, R.~F. 1995,
  \href{http://dx.doi.org/10.1086/513302}{\JournalTitle{\apjl}, 453, L49}

\bibitem[{{Ogilvie}(2014)}]{ogilvie14}
{Ogilvie}, G.~I. 2014,
  \href{http://dx.doi.org/10.1146/annurev-astro-081913-035941}{\JournalTitle{\araa},
  52, 171}

\bibitem[{{Ogilvie} \& {Lin}(2004)}]{ogilvielin07}
{Ogilvie}, G.~I., \& {Lin}, D.~N.~C. 2004,
  \href{http://dx.doi.org/10.1086/421454}{\JournalTitle{\apj}, 610, 477}

\bibitem[{{Ogilvie} \& {Lin}(2007)}]{ogilvie07}
---. 2007, \href{http://dx.doi.org/10.1086/515435}{\JournalTitle{\apj}, 661,
  1180}

\bibitem[{{Pall{\'e}} {et~al.}(2009){Pall{\'e}}, {Zapatero Osorio}, {Barrena},
  {Monta{\~n}{\'e}s-Rodr{\'{\i}}guez}, \& {Mart{\'{\i}}n}}]{palle09}
{Pall{\'e}}, E., {Zapatero Osorio}, M.~R., {Barrena}, R.,
  {Monta{\~n}{\'e}s-Rodr{\'{\i}}guez}, P., \& {Mart{\'{\i}}n}, E.~L. 2009,
  \href{http://dx.doi.org/10.1038/nature08050}{\JournalTitle{\nat}, 459, 814}

\bibitem[{{Parker}(1964)}]{parker64}
{Parker}, E.~N. 1964,
  \href{http://dx.doi.org/10.1086/147740}{\JournalTitle{\apj}, 139, 72}

\bibitem[{{Passy} {et~al.}(2012){Passy}, {Mac Low}, \& {De Marco}}]{passy12}
{Passy}, J.-C., {Mac Low}, M.-M., \& {De Marco}, O. 2012,
  \href{http://dx.doi.org/10.1088/2041-8205/759/2/L30}{\JournalTitle{\apjl},
  759, L30}

\bibitem[{{Peale} {et~al.}(1979){Peale}, {Cassen}, \& {Reynolds}}]{peale79}
{Peale}, S.~J., {Cassen}, P., \& {Reynolds}, R.~T. 1979,
  \href{http://dx.doi.org/10.1126/science.203.4383.892}{\JournalTitle{Science},
  203, 892}

\bibitem[{{Perez-Becker} \& {Chiang}(2013)}]{perez-becker13}
{Perez-Becker}, D., \& {Chiang}, E. 2013,
  \href{http://dx.doi.org/10.1093/mnras/stt895}{\JournalTitle{\mnras}, 433,
  2294}

\bibitem[{{Poppe}(2016)}]{poppe2016}
{Poppe}, A.~R. 2016,
  \href{http://dx.doi.org/10.1016/j.icarus.2015.10.001}{\JournalTitle{\icarus},
  264, 369}

\bibitem[{{Postberg} {et~al.}(2009){Postberg}, {Kempf}, {Schmidt},
  {Brilliantov}, {Beinsen}, {Abel}, {Buck}, \& {Srama}}]{postberg09}
{Postberg}, F., {Kempf}, S., {Schmidt}, J., {et~al.} 2009,
  \href{http://dx.doi.org/10.1038/nature08046}{\JournalTitle{\nat}, 459, 1098}

\bibitem[{{Potter} \& {Morgan}(1985)}]{pottermorgan85}
{Potter}, A., \& {Morgan}, T. 1985,
  \href{http://dx.doi.org/10.1126/science.229.4714.651}{\JournalTitle{Science},
  229, 651}

\bibitem[{{Potter} \& {Morgan}(1986)}]{pottermorgan86}
{Potter}, A.~E., \& {Morgan}, T.~H. 1986,
  \href{http://dx.doi.org/10.1016/0019-1035(86)90113-2}{\JournalTitle{\icarus},
  67, 336}

\bibitem[{{Potter} \& {Morgan}(1988)}]{pottermorgan88}
---. 1988,
  \href{http://dx.doi.org/10.1126/science.241.4866.675}{\JournalTitle{Science},
  241, 675}

\bibitem[{{Punzi} {et~al.}(2018){Punzi}, {Kastner}, {Melis}, {Zuckerman},
  {Pilachowski}, {Gingerich}, \& {Knapp}}]{punzi18}
{Punzi}, K.~M., {Kastner}, J.~H., {Melis}, C., {et~al.} 2018,
  \href{http://dx.doi.org/10.3847/1538-3881/aa9524}{\JournalTitle{\aj}, 155,
  33}

\bibitem[{{Rappaport} {et~al.}(2012){Rappaport}, {Levine}, {Chiang}, {El
  Mellah}, {Jenkins}, {Kalomeni}, {Kite}, {Kotson}, {Nelson},
  {Rousseau-Nepton}, \& {Tran}}]{rappaport2012}
{Rappaport}, S., {Levine}, A., {Chiang}, E., {et~al.} 2012,
  \href{http://dx.doi.org/10.1088/0004-637X/752/1/1}{\JournalTitle{\apj}, 752,
  1}

\bibitem[{Ridden-Harper {et~al.}(2016)}]{Ridden-Harper2016}
Ridden-Harper, A.~R., {et~al.} 2016, \JournalTitle{A\&A}, 593

\bibitem[{{Rogers}(2017)}]{Rogers17}
{Rogers}, T.~M. 2017,
  \href{http://dx.doi.org/10.1038/s41550-017-0131}{\JournalTitle{Nature
  Astronomy}, 1, 0131}

\bibitem[{{Rybicki} \& {Lightman}(1979)}]{rybickiandlightman}
{Rybicki}, G.~B., \& {Lightman}, A.~P. 1979, {Radiative processes in
  astrophysics}

\bibitem[{{Salz} {et~al.}(2018){Salz}, {Czesla}, {Schneider}, {Nagel},
  {Schmitt}, {Nortmann}, {Alonso-Floriano}, {L{\'o}pez-Puertas}, {Lamp{\'o}n},
  {Bauer}, {Snellen}, {Pall{\'e}}, {Caballero}, {Yan}, {Chen}, {Sanz- Forcada},
  {Amado}, {Quirrenbach}, {Ribas}, {Reiners}, {B{\'e}jar}, {Casasayas-Barris},
  {Cort{\'e}s-Contreras}, {Dreizler}, {Guenther}, {Henning}, {Jeffers},
  {Kaminski}, {K{\"u}rster}, {Lafarga}, {Lara}, {Molaverdikhani}, {Montes},
  {Morales}, {S{\'a}nchez-L{\'o}pez}, {Seifert}, {Zapatero Osorio}, \&
  {Zechmeister}}]{he2018}
{Salz}, M., {Czesla}, S., {Schneider}, P.~C., {et~al.} 2018,
  \href{http://dx.doi.org/10.1051/0004-6361/201833694}{\JournalTitle{\aap},
  620, A97}

\bibitem[{Schmidt(2016)}]{schmidt16comet}
Schmidt, C. 2016,
  \href{http://dx.doi.org/https://doi.org/10.1016/j.icarus.2015.10.009}{\JournalTitle{Icarus},
  265, 35 }

\bibitem[{{Schmidt} {et~al.}(2015{\natexlab{a}}){Schmidt}, {Johnson},
  {Mendillo}, {Baumgardner}, {Moore}, {O'Donoghue}, \&
  {Leblanc}}]{schmidt15AGU}
{Schmidt}, C., {Johnson}, R.~E., {Mendillo}, M., {et~al.} 2015{\natexlab{a}},
  \JournalTitle{AGU Fall Meeting Abstracts}, SM31B

\bibitem[{{Schmidt} {et~al.}(2016){Schmidt}, {Reardon}, {Killen}, {Gary},
  {Ahn}, {Leblanc}, {Baumgardner}, {Mendillo}, {Beck}, \&
  {Mangano}}]{schmidt16}
{Schmidt}, C., {Reardon}, K., {Killen}, R.~M., {et~al.} 2016, \JournalTitle{AGU
  Fall Meeting Abstracts}, P53B

\bibitem[{{Schmidt}(2013{\natexlab{a}})}]{schmidtPHD}
{Schmidt}, C.~A. 2013{\natexlab{a}}, PhD thesis, Boston University

\bibitem[{{Schmidt}(2013{\natexlab{b}})}]{schmidt13}
---. 2013{\natexlab{b}},
  \href{http://dx.doi.org/10.1002/jgra.50396}{\JournalTitle{Journal of
  Geophysical Research (Space Physics)}, 118, 4564}

\bibitem[{{Schmidt} {et~al.}(2015{\natexlab{b}}){Schmidt}, {Johnson},
  {Baumgardner}, \& {Mendillo}}]{schmidt15}
{Schmidt}, C.~A., {Johnson}, R.~E., {Baumgardner}, J., \& {Mendillo}, M.
  2015{\natexlab{b}},
  \href{http://dx.doi.org/10.1016/j.icarus.2014.10.022}{\JournalTitle{\icarus},
  247, 313}

\bibitem[{{Schmidt} {et~al.}(2018){Schmidt}, {Leblanc}, {Reardon}, {Killen},
  {Gary}, \& {Ahn}}]{schmidt18lpi}
{Schmidt}, C.~A., {Leblanc}, F., {Reardon}, K., {et~al.} 2018, in LPI
  Contributions, Vol. 2047, Mercury: Current and Future Science of the
  Innermost Planet, 6022

\bibitem[{{Schneider} {et~al.}(2009){Schneider}, {Burger}, {Schaller}, {Brown},
  {Johnson}, {Kargel}, {Dougherty}, \& {Achilleos}}]{schneider09}
{Schneider}, N.~M., {Burger}, M.~H., {Schaller}, E.~L., {et~al.} 2009,
  \href{http://dx.doi.org/10.1038/nature08070}{\JournalTitle{\nat}, 459, 1102}

\bibitem[{{Schneider} {et~al.}(1991){Schneider}, {Hunten}, {Wells}, {Schultz},
  \& {Fink}}]{schneider91}
{Schneider}, N.~M., {Hunten}, D.~M., {Wells}, W.~K., {Schultz}, A.~B., \&
  {Fink}, U. 1991, \href{http://dx.doi.org/10.1086/169694}{\JournalTitle{\apj},
  368, 298}

\bibitem[{{Seager} \& {Sasselov}(2000)}]{seagersasselov2000}
{Seager}, S., \& {Sasselov}, D.~D. 2000,
  \href{http://dx.doi.org/10.1086/309088}{\JournalTitle{\apj}, 537, 916}

\bibitem[{{Seidel} {et~al.}(2019){Seidel}, {Ehrenreich}, {Wyttenbach},
  {Allart}, {Lendl}, {Pino}, {Bourrier}, {Cegla}, {Lovis}, {Barrado},
  {Bayliss}, {Astudillo-Defru}, {Deline}, {Fisher}, {Heng}, {Joseph}, {Lavie},
  {Melo}, {Pepe}, {S{\'e}grasan}, \& {Udry}}]{Seidel19}
{Seidel}, J.~V., {Ehrenreich}, D., {Wyttenbach}, A., {et~al.} 2019,
  \JournalTitle{arXiv e-prints},
  \href{http://arxiv.org/abs/1902.00001}{{\sffamily arXiv:1902.00001
  [astro-ph.EP]}}

\bibitem[{{Sing} {et~al.}(2011){Sing}, {D{\'e}sert}, {Fortney}, {Lecavelier Des
  Etangs}, {Ballester}, {Cepa}, {Ehrenreich}, {L{\'o}pez-Morales}, {Pont},
  {Shabram}, \& {Vidal-Madjar}}]{sing11}
{Sing}, D.~K., {D{\'e}sert}, J.-M., {Fortney}, J.~J., {et~al.} 2011,
  \href{http://dx.doi.org/10.1051/0004-6361/201015579}{\JournalTitle{\aap},
  527, A73}

\bibitem[{{Sing} {et~al.}(2012){Sing}, {Huitson}, {Lopez-Morales}, {Pont},
  {D{\'e}sert}, {Ehrenreich}, {Wilson}, {Ballester}, {Fortney}, {Lecavelier des
  Etangs}, \& {Vidal-Madjar}}]{sing12_xo2b_na}
{Sing}, D.~K., {Huitson}, C.~M., {Lopez-Morales}, M., {et~al.} 2012,
  \href{http://dx.doi.org/10.1111/j.1365-2966.2012.21938.x}{\JournalTitle{\mnras},
  426, 1663}

\bibitem[{{Sing} {et~al.}(2015){Sing}, {Wakeford}, {Showman}, {Nikolov},
  {Fortney}, {Burrows}, {Ballester}, {Deming}, {Aigrain}, {D{\'e}sert},
  {Gibson}, {Henry}, {Knutson}, {Lecavelier des Etangs}, {Pont},
  {Vidal-Madjar}, {Williamson}, \& {Wilson}}]{sing15_w31b_K}
{Sing}, D.~K., {Wakeford}, H.~R., {Showman}, A.~P., {et~al.} 2015,
  \href{http://dx.doi.org/10.1093/mnras/stu2279}{\JournalTitle{\mnras}, 446,
  2428}

\bibitem[{Sing {et~al.}(2016)}]{Sing2016}
Sing, D.~K., {et~al.} 2016, \JournalTitle{Nature}, 529

\bibitem[{{Skrutskie} {et~al.}(2017){Skrutskie}, {de Kleer}, {Stone}, {Conrad},
  {Davies}, {de Pater}, {Leisenring}, {Hinz}, {Skemer}, {Veillet}, {Woodward},
  {Ertel}, \& {Spalding}}]{skrutskie17}
{Skrutskie}, M.~F., {de Kleer}, K.~R., {Stone}, J., {et~al.} 2017, in
  AAS/Division for Planetary Sciences Meeting Abstracts, Vol.~49, AAS/Division
  for Planetary Sciences Meeting Abstracts \#49, 407.02

\bibitem[{Snellen {et~al.}(2008)}]{Snellen2008}
Snellen, I. A.~G., {et~al.} 2008, \JournalTitle{A\&A}, 487

\bibitem[{{Spake} {et~al.}(2018){Spake}, {Sing}, {Evans}, {Oklop{\v c}i{\'c}},
  {Bourrier}, {Kreidberg}, {Rackham}, {Irwin}, {Ehrenreich}, {Wyttenbach},
  {Wakeford}, {Zhou}, {Chubb}, {Nikolov}, {Goyal}, {Henry}, {Williamson},
  {Blumenthal}, {Anderson}, {Hellier}, {Charbonneau}, {Udry}, \&
  {Madhusudhan}}]{spake2018}
{Spake}, J.~J., {Sing}, D.~K., {Evans}, T.~M., {et~al.} 2018,
  \href{http://dx.doi.org/10.1038/s41586-018-0067-5}{\JournalTitle{\nat}, 557,
  68}

\bibitem[{{Spencer} {et~al.}(2000){Spencer}, {Rathbun}, {Travis}, {Tamppari},
  {Barnard}, {Martin}, \& {McEwen}}]{spencer2000}
{Spencer}, J.~R., {Rathbun}, J.~A., {Travis}, L.~D., {et~al.} 2000,
  \href{http://dx.doi.org/10.1126/science.288.5469.1198}{\JournalTitle{Science},
  288, 1198}

\bibitem[{{Spencer} {et~al.}(2007){Spencer}, {Stern}, {Cheng}, {Weaver},
  {Reuter}, {Retherford}, {Lunsford}, {Moore}, {Abramov}, {Lopes}, {Perry},
  {Kamp}, {Showalter}, {Jessup}, {Marchis}, {Schenk}, \& {Dumas}}]{spencer07}
{Spencer}, J.~R., {Stern}, S.~A., {Cheng}, A.~F., {et~al.} 2007,
  \href{http://dx.doi.org/10.1126/science.1147621}{\JournalTitle{Science}, 318,
  240}

\bibitem[{{Sprague} {et~al.}(1997){Sprague}, {Kozlowski}, {Hunten},
  {Schneider}, {Domingue}, {Wells}, {Schmitt}, \& {Fink}}]{sprague97}
{Sprague}, A.~L., {Kozlowski}, R.~W.~H., {Hunten}, D.~M., {et~al.} 1997,
  \href{http://dx.doi.org/10.1006/icar.1997.5784}{\JournalTitle{\icarus}, 129,
  506}

\bibitem[{{Sudarsky} {et~al.}(2000){Sudarsky}, {Burrows}, \&
  {Pinto}}]{sudarsky2000}
{Sudarsky}, D., {Burrows}, A., \& {Pinto}, P. 2000,
  \href{http://dx.doi.org/10.1086/309160}{\JournalTitle{\apj}, 538, 885}

\bibitem[{{Sullivan} \& {Hunten}(1962)}]{sullivanhunten62}
{Sullivan}, H.~M., \& {Hunten}, D.~M. 1962,
  \href{http://dx.doi.org/10.1038/195589a0}{\JournalTitle{\nat}, 195, 589}

\bibitem[{{Sullivan} \& {Hunten}(1964)}]{sullivanhunten64}
---. 1964, \href{http://dx.doi.org/10.1139/p64-087}{\JournalTitle{Canadian
  Journal of Physics}, 42, 937}

\bibitem[{{Szalay} {et~al.}(2016){Szalay}, {Hor{\'a}nyi}, {Colaprete}, \&
  {Sarantos}}]{szalay16}
{Szalay}, J.~R., {Hor{\'a}nyi}, M., {Colaprete}, A., \& {Sarantos}, M. 2016,
  \href{http://dx.doi.org/10.1002/2016GL069541}{\JournalTitle{\grl}, 43, 6096}

\bibitem[{{Teachey} \& {Kipping}(2018)}]{teacheykipping18}
{Teachey}, A., \& {Kipping}, D.~M. 2018,
  \href{http://dx.doi.org/10.1126/sciadv.aav1784}{\JournalTitle{Science
  Advances}, 4, eaav1784}

\bibitem[{{Thomas}(1996)}]{thomas96}
{Thomas}, N. 1996, \JournalTitle{\aap}, 313, 306

\bibitem[{{Thomas} {et~al.}(2004){Thomas}, {Bagenal}, {Hill}, \&
  {Wilson}}]{thomas04}
{Thomas}, N., {Bagenal}, F., {Hill}, T.~W., \& {Wilson}, J.~K. 2004, {The Io
  neutral clouds and plasma torus}, ed. F.~{Bagenal}, T.~E. {Dowling}, \& W.~B.
  {McKinnon}, 561

\bibitem[{{Thorngren} {et~al.}(2016){Thorngren}, {Fortney}, {Murray-Clay}, \&
  {Lopez}}]{thorngren2016}
{Thorngren}, D.~P., {Fortney}, J.~J., {Murray-Clay}, R.~A., \& {Lopez}, E.~D.
  2016,
  \href{http://dx.doi.org/10.3847/0004-637X/831/1/64}{\JournalTitle{\apj}, 831,
  64}

\bibitem[{{Trafton}(1975)}]{trafton75}
{Trafton}, L. 1975,
  \href{http://dx.doi.org/10.1038/258690a0}{\JournalTitle{\nat}, 258, 690}

\bibitem[{{Tsang} {et~al.}(2016){Tsang}, {Spencer}, {Lellouch},
  {Lopez-Valverde}, \& {Richter}}]{tsang16}
{Tsang}, C.~C.~C., {Spencer}, J.~R., {Lellouch}, E., {Lopez-Valverde}, M.~A.,
  \& {Richter}, M.~J. 2016,
  \href{http://dx.doi.org/10.1002/2016JE005025}{\JournalTitle{Journal of
  Geophysical Research (Planets)}, 121, 1400}

\bibitem[{{Tsang} {et~al.}(2013){Tsang}, {Spencer}, {Lellouch},
  {L{\'o}pez-Valverde}, {Richter}, {Greathouse}, \& {Roe}}]{tsang13}
{Tsang}, C. C.~C., {Spencer}, J.~R., {Lellouch}, E., {et~al.} 2013,
  \href{http://dx.doi.org/10.1016/j.icarus.2013.06.032}{\JournalTitle{\icarus},
  226, 1177}

\bibitem[{{Valek} {et~al.}(2018){Valek}, {McComas}, {Kurth}, {Hospodarsky},
  {Thomsen}, {Levin}, {Wilson}, {Mauk}, {Ebert}, {Bolton}, {Gladstone},
  {Louarn}, {Bagenal}, {Allegrini}, {Greathouse}, {Connerney}, {Clark}, {Hue},
  \& {Pollock}}]{valek18}
{Valek}, P., {McComas}, D., {Kurth}, W., {et~al.} 2018, in COSPAR Meeting,
  Vol.~42, 42nd COSPAR Scientific Assembly, B5.1

\bibitem[{{van Lieshout} {et~al.}(2014){van Lieshout}, {Min}, \&
  {Dominik}}]{vanlieshout14}
{van Lieshout}, R., {Min}, M., \& {Dominik}, C. 2014,
  \href{http://dx.doi.org/10.1051/0004-6361/201424876}{\JournalTitle{\aap},
  572, A76}

\bibitem[{{Vanderburg} {et~al.}(2015){Vanderburg}, {Johnson}, {Rappaport},
  {Bieryla}, {Irwin}, {Lewis}, {Kipping}, {Brown}, {Dufour}, {Ciardi}, {Angus},
  {Schaefer}, {Latham}, {Charbonneau}, {Beichman}, {Eastman}, {McCrady},
  {Wittenmyer}, \& {Wright}}]{vanderburg15}
{Vanderburg}, A., {Johnson}, J.~A., {Rappaport}, S., {et~al.} 2015,
  \href{http://dx.doi.org/10.1038/nature15527}{\JournalTitle{\nat}, 526, 546}

\bibitem[{Vidal-Madjar {et~al.}(2011)}]{Vidal-Madjar2011}
Vidal-Madjar, A., {et~al.} 2011, \JournalTitle{A\&A}, 527

\bibitem[{{Vidal-Madjar} {et~al.}(2013){Vidal-Madjar}, {Huitson}, {Bourrier},
  {D{\'e}sert}, {Ballester}, {Lecavelier des Etangs}, {Sing}, {Ehrenreich},
  {Ferlet}, {H{\'e}brard}, \& {McConnell}}]{mgvidal13}
{Vidal-Madjar}, A., {Huitson}, C.~M., {Bourrier}, V., {et~al.} 2013,
  \href{http://dx.doi.org/10.1051/0004-6361/201322234}{\JournalTitle{\aap},
  560, A54}

\bibitem[{{Volkov} \& {Johnson}(2013)}]{volkovjohnson13}
{Volkov}, A.~N., \& {Johnson}, R.~E. 2013,
  \href{http://dx.doi.org/10.1088/0004-637X/765/2/90}{\JournalTitle{\apj}, 765,
  90}

\bibitem[{{Volkov} {et~al.}(2011){Volkov}, {Johnson}, {Tucker}, \&
  {Erwin}}]{volkov11}
{Volkov}, A.~N., {Johnson}, R.~E., {Tucker}, O.~J., \& {Erwin}, J.~T. 2011,
  \href{http://dx.doi.org/10.1088/2041-8205/729/2/L24}{\JournalTitle{\apjl},
  729, L24}

\bibitem[{{Watson} {et~al.}(1981){Watson}, {Donahue}, \& {Walker}}]{watson81}
{Watson}, A.~J., {Donahue}, T.~M., \& {Walker}, J.~C.~G. 1981,
  \href{http://dx.doi.org/10.1016/0019-1035(81)90101-9}{\JournalTitle{\icarus},
  48, 150}

\bibitem[{{Weidner} \& {Horne}(2010)}]{weidnerhorne10}
{Weidner}, C., \& {Horne}, K. 2010,
  \href{http://dx.doi.org/10.1051/0004-6361/201014955}{\JournalTitle{\aap},
  521, A76}

\bibitem[{{Wilson} {et~al.}(2002){Wilson}, {Mendillo}, {Baumgardner},
  {Schneider}, {Trauger}, \& {Flynn}}]{wilson02}
{Wilson}, J.~K., {Mendillo}, M., {Baumgardner}, J., {et~al.} 2002,
  \href{http://dx.doi.org/10.1006/icar.2002.6821}{\JournalTitle{\icarus}, 157,
  476}

\bibitem[{{Wilson} {et~al.}(2006){Wilson}, {Mendillo}, \& {Spence}}]{wilson06}
{Wilson}, J.~K., {Mendillo}, M., \& {Spence}, H.~E. 2006,
  \href{http://dx.doi.org/10.1029/2005JA011364}{\JournalTitle{Journal of
  Geophysical Research (Space Physics)}, 111, A07207}

\bibitem[{Wilson {et~al.}(2015)}]{Wilson2015}
Wilson, P.~A., {et~al.} 2015, \JournalTitle{MNRAS}, 450

\bibitem[{{Wong} \& {Johnson}(1996{\natexlab{a}})}]{wongjohnson96a}
{Wong}, M.~C., \& {Johnson}, R.~E. 1996{\natexlab{a}},
  \href{http://dx.doi.org/10.1029/96JE02510}{\JournalTitle{\jgr}, 101, 23243}

\bibitem[{{Wong} \& {Johnson}(1996{\natexlab{b}})}]{wongjohnson96b}
---. 1996{\natexlab{b}},
  \href{http://dx.doi.org/10.1029/96JE02509}{\JournalTitle{\jgr}, 101, 23255}

\bibitem[{{Wu}(2005{\natexlab{a}})}]{wu05-1}
{Wu}, Y. 2005{\natexlab{a}},
  \href{http://dx.doi.org/10.1086/497354}{\JournalTitle{\apj}, 635, 674}

\bibitem[{{Wu}(2005{\natexlab{b}})}]{wu05-2}
---. 2005{\natexlab{b}},
  \href{http://dx.doi.org/10.1086/497355}{\JournalTitle{\apj}, 635, 688}

\bibitem[{{Wyttenbach} {et~al.}(2015){Wyttenbach}, {Ehrenreich}, {Lovis},
  {Udry}, \& {Pepe}}]{wyttenbach15}
{Wyttenbach}, A., {Ehrenreich}, D., {Lovis}, C., {Udry}, S., \& {Pepe}, F.
  2015,
  \href{http://dx.doi.org/10.1051/0004-6361/201525729}{\JournalTitle{\aap},
  577, A62}

\bibitem[{{Wyttenbach} {et~al.}(2017){Wyttenbach}, {Lovis}, {Ehrenreich},
  {Bourrier}, {Pino}, {Allart}, {Astudillo-Defru}, {Cegla}, {Heng}, {Lavie},
  {Melo}, {Murgas}, {Santerne}, {S{\'e}gransan}, {Udry}, \&
  {Pepe}}]{wyttenbach17}
{Wyttenbach}, A., {Lovis}, C., {Ehrenreich}, D., {et~al.} 2017,
  \href{http://dx.doi.org/10.1051/0004-6361/201630063}{\JournalTitle{\aap},
  602, A36}

\bibitem[{{Zuluaga} {et~al.}(2015){Zuluaga}, {Kipping}, {Sucerquia}, \&
  {Alvarado}}]{exorings}
{Zuluaga}, J.~I., {Kipping}, D.~M., {Sucerquia}, M., \& {Alvarado}, J.~A. 2015,
  \href{http://dx.doi.org/10.1088/2041-8205/803/1/L14}{\JournalTitle{\apjl},
  803, L14}

\end{thebibliography}

\end{document}